\documentclass[12pt, draftclsnofoot, onecolumn]{IEEEtran}
\usepackage{amsmath,amsfonts}
\usepackage{algorithmic}
\usepackage{algorithm}
\usepackage{array}
\usepackage {subfigure}
\usepackage[caption=false,font=normalsize,labelfont=sf,textfont=sf]{subfig}
\usepackage{textcomp}
\usepackage{stfloats}
\usepackage{url}
\usepackage{verbatim}
\usepackage{graphicx}
\usepackage{cite}
\usepackage{amssymb}
\usepackage{stfloats}
\usepackage{cuted}
\begin{document}
\title{RIS-Enhanced Cognitive Integrated Sensing and Communication: Joint Beamforming \\and Spectrum Sensing}

\author{Yongqing Xu, Yong Li, \emph{Member, IEEE}, and Tony Q. S. Quek, \emph{Fellow, IEEE}\\
	
	% <-this % stops a space
	% \thanks{\emph{(Corresponding author: Yong Li)}}% <-this % stops a space
	\thanks{The work of Y. Xu and Y. Li was supported by the National Research and Development Program ``Distributed Large Dimensional Wireless Cooperative Transmission Technology Research and System Verification'' under Grant 2022YFB2902400. The work of Tony Q. S. Quek was supported by the National Research Foundation, Singapore and Infocomm Media Development Authority under its Future Communications Research \& Development Programme. (Corresponding Author: Yong Li and Tony Q. S. Quek.)
		
		Y. Xu and Y. Li are with the Key Laboratory of Universal Wireless Communications, Beijing University of Posts and Telecommunications, Beijing 100876, China. Email: \{xuyongqing; liyong\}@bupt.edu.cn. 
		
		T. Q. S. Quek is with the Singapore University of Technology and Design, Singapore 487372, and also with the Yonsei Frontier Lab, Yonsei University, South Korea (e-mail: tonyquek@sutd.edu.sg).
	}}%



\maketitle
\thispagestyle{empty}
\begin{abstract}
Cognitive radio (CR) and integrated sensing and communication (ISAC) are both critical technologies for the sixth generation (6G) wireless networks. However, their interplay has yet to be explored. To obtain the mutual benefits between CR and ISAC, we focus on a reconfigurable intelligent surface (RIS)-enhanced cognitive ISAC system and explore using the additional degrees-of-freedom brought by the RIS to improve the performance of the cognitive ISAC system. Specifically, we formulate an optimization problem of maximizing the signal-to-noise-plus-interference ratios (SINRs) of the mobile sensors (MSs) while ensuring the requirements of the spectrum sensing (SS) and the secondary transmissions by jointly designing the SS time, the secondary base station (SBS) beamforming, and the RIS beamforming. The formulated non-convex problem can be solved by the proposed block coordinate descent (BCD) algorithm based on the Dinkelbach's transform and the successive convex approximation (SCA) methods. Simulation results demonstrate that the proposed scheme exhibits good convergence performance and can effectively reduce the position error bounds (PEBs) of the MSs, thereby improving the radio environment map (REM) accuracy of CR networks. Additionally, we reveal the impact of RIS deployment locations on the performance of cognitive ISAC systems.
\end{abstract}

\addtolength{\topmargin}{-.05in}
\section{Introduction}
\IEEEPARstart{M}{any} key technologies have emerged with the development of the research on the sixth generation (6G) wireless communication systems, such as integrated sensing and communication (ISAC), reconfigurable intelligent surface (RIS), and ultra-massive multiple-input multiple-output (MIMO) \cite{wang2023road}. The intersections between these emerging technologies and the existing technologies, such as the cognitive radio (CR), (e.g., RIS and ISAC \cite{xu2023joint,zhang2022active,wang2023stars,chepuri2023integrated}, RIS and CR \cite{wu2021irs,makarfi2021reconfigurable,nasser2022intelligent,he2020reconfigurable,yuan2020intelligent,xu2020resource,wu2023joint,deng2023joint}) have the potentials to further enhance the system performance of 6G. For example, on the one hand, the CR can further enhance the spectral efficiency of the ISAC systems; on the other hand, the ISAC can improve the capacity of the spectrum sensing (SS) for the CR networks.

CR technology enables the secondary nodes to access the unused licensed spectrum opportunistically and hence improves the utilization efficiency of spectrum \cite{ahmad2015survey}. The CR network faces the hidden node problem, which is caused by the secondary nodes being unable to detect the transmissions of the primary nodes due to the impact of severe path loss. A radio environment map (REM), which includes spectrum usage information at any location in the CR network, has emerged to address the hidden node problem \cite{yilmaz2013radio}. The REM is constructed by interpolating the spectrum usage information gathered by the secondary nodes with known positions based on the spatial correlation. Therefore, the accuracy of REM mainly depends on the interpolation error and the localization errors of the secondary nodes. For example, the authors of \cite{sato2017kriging} analyzed the statistical characteristic of the interpolation error and then formulated a closed-form expression for the maximum transmit power of the secondary node. The authors of \cite{augusto2018geostatistical} used a geostatistical prediction tool to improve the REM accuracy under the localization error condition. The authors of \cite{zhen2022radio} proposed a machine learning algorithm to deal with the positional uncertainty when constructing a REM. Nevertheless, the accuracy of REM will be severely compromised when the localization error of the secondary nodes is large.

Research on ISAC is currently underway to enable 6G wireless systems to simultaneously realize communication and target sensing \cite{zhang2021enabling}. The accuracy of REM can be improved by using the ISAC to reduce the localization error. Specifically, in \cite{chiriyath2015inner} and \cite{xiong2022flowing}, the authors evaluated the tradeoff between sensing and communication in ISAC systems using the proposed performance metrics, such as the radar estimation information rate, the Fisher information, and the bound of communication rate. In \cite{tang2010mimo}, the authors proposed to use the radar mutual information to design radar waveforms. Moreover, in \cite{liu2018mu}, the authors used the convex optimization and the Riemannian manifold optimization methods to design the ISAC waveforms under the separated and shared antenna deployments. In \cite{tian2021transmit}, the authors designed the transmit/receive radar beam patterns and the receive communication beam pattern to maximize the Kullback-Leibler divergence (KLD) while satisfying the signal-to-noise-plus-interference (SINR) constraints of communication users. In \cite{dong2022sensing}, the authors proposed the concept of sensing quality of service (QoS) and a unified resource allocation framework to satisfy different sensing QoSs. However, joint design of communication and sensing limits the degrees-of-freedom (DoFs) when designing the ISAC waveforms.

A RIS is a planar surface composed of numerous low-cost and nearly passive reconﬁgurable elements that dynamically induce appropriate amplitude and phase shifts to the incident signals \cite{di2020smart,liu2021reconfigurable,hou2020reconfigurable}. By leveraging the RIS technology, the DoFs for designing ISAC waveforms can be increased. For example, the authors of \cite{xu2023joint,zhang2022active,wang2023stars,chepuri2023integrated} designed the joint BS and RIS beamforming to satisfy the communication and sensing requirements simultaneously. Moreover, RIS has been shown to enhance the capacity of secondary nodes to detect the primary signals in \cite{wu2021irs} and \cite{makarfi2021reconfigurable}, i.e., the capacity of the SS. The authors of \cite{wu2021irs} investigated using RISs to enhance the energy detection for SS and derived the closed-form expressions for the probability of detection. The authors of \cite{makarfi2021reconfigurable} considered two RIS configurations, i.e., the access point RIS and the RIS relay, and analyzed the performance of the energy detection. 

Furthermore, in \cite{nasser2022intelligent,he2020reconfigurable,yuan2020intelligent,xu2020resource}, all the authors considered using RISs to increase the communication rates of the secondary networks while satisfying the interference tolerance of the primary networks. More importantly, the authors of \cite{wu2023joint} and \cite{deng2023joint} used RISs to simultaneously enhance the capacity of SS and the performance of the secondary transmission by striking a balance between the SS and the secondary transmission. The authors of \cite{liang2023extending} utilized the spectrum usage information to dynamically adjust the transmitter parameters to satisfy the communication rate and target detection probability requirements. To the best of our knowledge, no study considers using RISs to enhance the capacities of the SS and the ISAC simultaneously or investigates the interplay between the CR and the ISAC.

\subsection{The Contributions of This Work}
In this paper, we aim to use a RIS to improve the localization accuracy of the secondary nodes, i.e., mobile sensors (MSs), for the sake of improving the accuracy of the REM while satisfying the performance of the SS, the communication rates of the secondary user equipments (SUEs), and the interference power of the primary user equipments (PUEs). The main contributions of our work are listed as follows:
\begin{itemize}
	\item 
	We establish signal models for the RIS-enhanced cognitive ISAC systems, including the SS signals of the secondary base station (SBS) and the received signals of the MSs and the SUEs, based on which the weighted position error bound (PEB) and the weighted SINR of the MS are derived.
	\item 
	We formulate an optimization problem for maximizing the SINRs of the MSs under the SS constraint, the constraints of the communication rates of the SUEs, and the constraints of the interference power of the PUEs. A block coordinate descent (BCD) algorithm based on the Dinkelbach’s transform and the successive convex approximation (SCA) methods are proposed to solve the formulated non-convex problem.
	\item
	We analyze the complexity and the convergence performance of the proposed algorithms. Moreover, we perform numerous simulations to verify the effectiveness of the proposed algorithms in various aspects.
	\item
	Specifically, we demonstrate that maximizing the SINRs of the MSs can effectively reduce the PEBs of the MSs and deploying the RIS at an arbitrary location does not necessarily guarantee the performance gain. 
\end{itemize}
\addtolength{\topmargin}{-.05in}
\subsection{Paper Organization}
The remainder of this paper is organized as follows. Section \uppercase\expandafter{\romannumeral2} introduces the system model and performance metrics. Section \uppercase\expandafter{\romannumeral3} proposes a BCD-based algorithm to design the sensing time and the joint beamforming. Section \uppercase\expandafter{\romannumeral4} analyzes the complexity and convergence performance of the proposed algorithms. Section \uppercase\expandafter{\romannumeral5} presents the simulation results. Section \uppercase\expandafter{\romannumeral6} concludes the paper.

\subsection{Notations}
Throughout this paper, matrices are denoted by bold uppercase letters, e.g., $\boldsymbol{X}$, vectors are denoted by bold lowercase letters, e.g., $\boldsymbol{x}$, and scalars are denoted by normal fonts, e.g., $x$.  $\left(\cdot\right)^T$, $\left(\cdot\right)^H$, and $\left(\cdot\right)^*$ represent the transpose, the conjugate transpose, and the conjugate, respectively. $Q(\cdot)$ represents the Q function. $\text{Tr}\left(\cdot\right)$ represents the trace operation. $\Im\left(\cdot\right)$ and $\Re\left(\cdot\right)$ denotes the real part and the imaginary part retrieval operations. $\frac{\partial(\cdot)}{\partial(\cdot)}$ represents the partial derivative operation. $\vert\cdot\vert$ denotes the absolute value of a one-dimensional number. $\Vert\cdot\Vert_2$ stands for the $l_2$ norm. $\text{Pr}\{\cdot\}$ represents the probability. $\text{diag}\left(\boldsymbol{x}\right)$ represents using $\boldsymbol{x}$ to formulate a diagonal matrix. $\text{diag}\left(\boldsymbol{X}\right)$ represents using the diagonal elements of $\boldsymbol{X}$ to formulate a column vector. $\text{rank}\left(\boldsymbol{X}\right)$ denotes the rank of $\boldsymbol{X}$.

\section{System Model}
\begin{figure}[!t]
	\centering
	\includegraphics[width=0.4\textwidth]{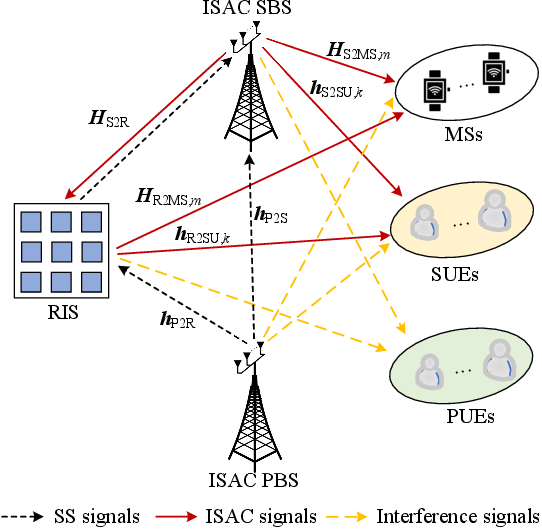}
	\caption{Considered RIS-enhanced cognitive ISAC system.}
	\label{fig1}
\end{figure}
The considered system is depicted in Fig. 1, where an ISAC SBS with $N_{\text{B}}$ antennas detects the signals, i.e., the SS signals, from a single-antenna ISAC  primary base station (PBS). The ISAC SBS will send the downlink ISAC signals to simultaneously localize $M$  MSs with each MS equipping $N_{\text{M}}$ antennas and communicate with $K$ single-antenna SUEs when it does not detect the presence of the signals from the ISAC PBS. The detection of the SS signals and the downlink transmission are both assisted by a RIS with $N_{\text{R}}$ elements. Two types of interference will exist when the ISAC SBS fails to detect the presence of the ISAC PBS signals. The first one is the interference signals from the ISAC SBS to $L$ PUEs, and the second one is from the PBS to the MSs and the SUEs. For clarity, the remarks on the interference signal are omitted in Fig. 1.
Moreover, the antennas and elements of the ISAC SBS, the MSs, and the RIS are all configured as uniform linear arrays (ULAs). The inter-distance of all antennas and RIS elements are half-wavelength. The sets of SUEs, PUEs, MSs, and RIS elements are represented by $\mathcal{K}$, $\mathcal{L}$, $\mathcal{M}$, and $\mathcal{N}$, respectively.

The channel vectors between the ISAC SBS and the ISAC PBS, between the SBS and the $k$-th SUE, and between the SBS and the $l$-th PUE are respectively denoted by $\boldsymbol{h}_{\text{S2P}}\in\mathbb{C}^{N_\text{B}\times1}$, $\boldsymbol{h}_{\text{S2SU},k}\in\mathbb{C}^{N_\text{B}\times1}$, and $\boldsymbol{h}_{\text{S2PU},l}\in\mathbb{C}^{N_\text{B}\times1}$; the channel matrices between the SBS and the RIS and between the SBS and the $m$-th MS are respectively denoted by $\boldsymbol{H}_{\text{S2R}}\in\mathbb{C}^{N_\text{B}\times N_\text{R}}$ and $\boldsymbol{H}_{\text{S2MS},m}\in\mathbb{C}^{N_\text{B}\times N_\text{M}}$; the channels between the ISAC PBS and the RIS, between the PBS and the $m$-th MS, and between the PBS and the $k$-th SUE are respectively denoted by $\boldsymbol{h}_{\text{P2R}}\in\mathbb{C}^{N_\text{R}\times1}$, $\boldsymbol{h}_{\text{P2MS},m}\in\mathbb{C}^{N_\text{M}\times1}$, $h_{\text{P2SU},k}$; the channels between the RIS and the $m$-th MS, between the RIS and the $k$-th SUE, and between the RIS and the $l$-th PUE are respectively denoted by $\boldsymbol{H}_{\text{R2MS},m}\in\mathbb{C}^{N_\text{R}\times N_\text{M}}$, $\boldsymbol{h}_{\text{R2SU},k}\in\mathbb{C}^{N_\text{R}\times1}$, and $\boldsymbol{h}_{\text{R2PU},l}\in\mathbb{C}^{N_\text{R}\times1}$.

We assume that all channels are perfectly estimated. We can use the methods in \cite{wang2020channel} to estimate the direct channels and the cascaded channels of RIS. Nevertheless, it is challenging to obtain the perfect channel state information (CSI) in some practical systems. We can use the methods in \cite{xu2022weighted} to mitigate the impact of the CSI errors. To focus on the main points of the paper, we will not delve into extensive discussions on the imperfect CSI.

\subsection{Signal Model for Spectrum Sensing}
We assume that the total downlink time duration is $T$, the SS time is $\tau$, and thus the time duration of the downlink signal transmission is $T-\tau$. The energy detection method is used to detect the signals from the PBS. The received signals of the SBS at the time instant $t$ can be expressed as
\begin{equation}
	\label{eq1}
	\begin{aligned}
		&\mathcal{H}_1:\boldsymbol{y}_{\text{SBS}}(t)\!=\!\sqrt{P_{\text{PBS}}}\left(\boldsymbol{h}_{\text{P2S}}\!+\!\boldsymbol{H}_{\text{S2R}}\boldsymbol{\Phi}\boldsymbol{h}_{\text{P2R}}\right)x_{\text{PBS}}(t)\!+\!\boldsymbol{n}_{\text{SBS}}(t),\\
		&\mathcal{H}_0:\boldsymbol{y}_{\text{SBS}}(t)=\boldsymbol{n}_{\text{SBS}}(t),
	\end{aligned}
\end{equation}
where $\mathcal{H}_1$ denotes the presence of the signals of the PBS, $\mathcal{H}_0$ denotes the absence of the PBS signals, $\sqrt{P_{\text{PBS}}}$ is the transmit power of the PBS, and $\boldsymbol{\Phi}\in\mathbb{C}^{N_{\text{R}}\times N_{\text{R}}}$ is the reflecting matrix of the RIS, which can be written as $\text{diag}\left(\left[\phi_1,\cdots,\phi_{N_{\text{R}}}\right]^T\right)$. We assume that the reflection magnitude of each RIS element equals one and the phase shift is continuous, i.e., $\left\vert\phi_n\right\vert=1,\forall n\in\mathcal{N}$. Moreover, $\boldsymbol{n}_{\text{SBS}}(t)$ is the additive white Gaussian noise (AWGN) with the variance $\sigma_{\text{SBS}}^2$. Then, the detection probability and the probability of false alarm are respectively defined as \cite{choi2013optimal}
\begin{equation}
	\label{eq2}
	P_d\left(\tau,\boldsymbol{\Phi}\right)\triangleq Q\left(\frac{\epsilon-\sigma_{\text{SBS}}^2\left(N_{\text{B}}+\text{SNR}_{\text{SBS}}\right)}{\left(\sigma_{\text{SBS}}^2/\sqrt{\tau f_s}\right)\sqrt{2\text{SNR}_{\text{SBS}}+N_{\text{B}}}}\right),
\end{equation}
and
\begin{equation}
	\label{eq3}
	P_f\left(\tau\right)\triangleq Q\left(\frac{\epsilon-\sigma_{\text{SBS}}^2N_{\text{B}}}{\left(\sigma_{\text{SBS}}^2/\sqrt{\tau f_s}\right)\sqrt{N_{\text{B}}}}\right),
\end{equation}
where $\text{SNR}_{\text{SBS}}$ is the signal-to-noise ratio (SNR) of the received signals of the SBS, $\text{SNR}_{\text{SBS}}\triangleq P_{\text{PBS}}\left\Vert\boldsymbol{h}_{\text{P2S}}+\boldsymbol{H}_{\text{S2R}}\boldsymbol{\Phi}\boldsymbol{h}_{\text{P2R}}\right\Vert_2^2/\sigma_{\text{SBS}}^2$, $\epsilon$ is the detection threshold of the energy detection method, and $f_s$ is the sampling frequency of the SBS. The SBS will transmit the downlink signals in two cases. The first case (Case 1) refers to when $\mathcal{H}_0$ is true and no false alarm occurs; the second case (Case 2) refers to when $\mathcal{H}_1$ is true and the missed detection occurs.

\subsection{Position Error Bound and Average Sensing SINR}
The received signals of the $m$-th MS can be expressed as
\begin{equation}
	\label{eq4}
	\begin{aligned}
	&\boldsymbol{y}_{\text{MS},m}(t)=\left(\boldsymbol{H}_{\text{R2MS},m}^T\boldsymbol{\Phi}\boldsymbol{H}_{\text{S2R}}^T+\boldsymbol{H}_{\text{S2MS},m}^T\right)\boldsymbol{w}_mx_m(t)\\
	&+\sum_{i=1,i\neq m}^{M+K}\left(\boldsymbol{H}_{\text{R2MS},m}^T\boldsymbol{\Phi}\boldsymbol{H}_{\text{S2R}}^T+\boldsymbol{H}_{\text{S2MS},m}^T\right)\boldsymbol{w}_ix_i(t)\\
	&+\underbrace{\sqrt{P_{\text{PBS}}}\left(\boldsymbol{h}_{\text{P2MS},m}+\boldsymbol{H}_{\text{R2MS},m}^T\boldsymbol{\Phi}\boldsymbol{h}_{\text{P2R}}\right)x_{\text{PBS}}(t)}_{\text{Case 2}}+\boldsymbol{n}_{\text{MS}}(t),
	\end{aligned}
\end{equation}
where $\boldsymbol{W}=\left[\boldsymbol{w}_1,\cdots,\boldsymbol{w}_M,\boldsymbol{w}_{M+1},\cdots,\boldsymbol{w}_{M+K}\right]\in\mathbb{C}^{N_{\text{B}}\times(M+K)}$ denotes the matrix of the transmit beamforming of SBS and $\boldsymbol{n}_{\text{MS}}(t)$ is the AWGN of the MS with the variance $\sigma_{\text{MS}}^2$. We assume that the orthogonal waveform is adopted to localized the MSs, i.e., $x_m(t),\forall m\in\mathcal{M}$ is assumed to be orthogonal to each other. Moreover, it is noted that the third term in the right-hand side of (\ref{eq4}) only exists when Case 2 introduced in the last subsection occurs.

The PEB of the $m$-th MS is defined as \cite{kay1993fundamentals}
\begin{equation}
	\label{eq5}
	\text{PEB}_{m}\triangleq\sqrt{\text{Tr}\left[\left(\boldsymbol{F}_{\text{p},m}^{-1}\right)_{1:2,1:2}\right]},
\end{equation}
where $\boldsymbol{F}_{\text{p},m}\in\mathbb{R}^{6\times6}$ respectively denotes the Fisher information matrix (FIM) of the position parameters of the $m$-th MS. The vector of the position parameters can be written as
\begin{equation}
	\label{eq6}
	\boldsymbol{\xi}_{\text{p},m}=\left[x_m,y_m,\Im\left\{h_{\text{D},m}\right\},\Re\left\{h_{\text{D},m}\right\},\Im\left\{h_{\text{R},m}\right\},\Re\left\{h_{\text{R},m}\right\}\right]^T\!\!\!\!,
\end{equation}
where $(x_m,y_m)$ represents the 2D position of the $m$-th MS, and $h_{\text{D},m}$ and $h_{\text{R},m}$ denote the large scale gains of the direct and the reflected channels between the SBS and the MS. For the convenience of deriving the FIM, we rewrite the direct and the reflected channel matrices between the SBS and the $m$-th MS as
\begin{equation}
	\label{eq7}
	\boldsymbol{H}_{\text{S2MS},m}=h_{\text{D}}\boldsymbol{a}\left(\theta_{\text{D}}^{\prime}\right)\boldsymbol{b}^T\left(\theta_{\text{D}}\right)
\end{equation}
and
\begin{equation}
	\label{eq8}
	\boldsymbol{H}_{\text{S2R}}\boldsymbol{\Phi}\boldsymbol{H}_{\text{R2MS},m}=h_{\text{R}}\boldsymbol{a}\left(\theta_{\text{R}}^{\prime}\right)\boldsymbol{c}^T\left(\theta_{\text{R}}^{\prime}\right)\boldsymbol{\Phi}\boldsymbol{c}\left(\theta_{\text{R}}\right)\boldsymbol{b}^T\left(\theta_{\text{R}}\right),
\end{equation}
where we ignore the subscripts on the right-hand side of the above two equations for clarify, such as S2MS and $m$. Moreover, $\theta_{\text{D}}^{\prime}$ and $\theta_{\text{R}}^{\prime}$ respectively denote the angles of departure (AoDs) of the direct and the reflected paths, $\theta_{\text{D}}$ and $\theta_{\text{R}}$ respectively denote the angles of arrival (AoAs) of the direct and the reflected paths, and $\boldsymbol{a}\left(\theta\right)$, $\boldsymbol{b}\left(\theta\right)$, and $\boldsymbol{c}\left(\theta\right)$ are the steering vectors of the SBS, the MS, and the RIS, respectively. All the steering vectors can be written as $\left[1,e^{\left(-j\pi\sin\theta\right)},\cdots,e^{\left(-j\pi(N-1)\sin\theta\right)}\right]^T$, where $N$ is the number of antenna elements. Since the received signals are not directlly related to the position parameters, we use the chain rule \cite{kay1993fundamentals} to rewrite the $\boldsymbol{F}_{\text{p}}$ as 
\begin{equation}
	\label{eq9}
	\boldsymbol{F}_{\text{p}}=\boldsymbol{\Gamma}^T\boldsymbol{F}_{\text{c}}\boldsymbol{\Gamma},
\end{equation}
where $\boldsymbol{\Gamma}\in\mathbb{R}^{6\times6}$ is the Jacobian matrix with $\boldsymbol{\Gamma}_{i,j}=\frac{\partial\left(\boldsymbol{\xi}_{\text{c}}\right)_i}{\partial\left(\boldsymbol{\xi}_{\text{p}}\right)_j}$, the vector of the channel parameters can be written as
\begin{equation}
	\label{eq10}
	\boldsymbol{\xi}_{\text{c}}=\left[\theta_{\text{D}},\theta_{\text{R}},\Im\left\{h_{\text{D}}\right\},\Re\left\{h_{\text{D}}\right\},\Im\left\{h_{\text{R}}\right\},\Re\left\{h_{\text{R}}\right\}\right]^T,
\end{equation}
and $\boldsymbol{F}_{\text{c}}$ denotes the FIM of the channel parameters, which is defined as
\begin{equation}
	\label{eq11}
	\boldsymbol{F}_{\text{c}}\triangleq\frac{2(T-\tau)}{\sigma^2}\sum_{n=1}^{N_{\text{M}}}\Re\left\{\frac{\partial\left(\overline{\boldsymbol{y}}\right)_n}{\partial\boldsymbol{\xi}_{\text{c}}}\left[\frac{\partial\left(\overline{\boldsymbol{y}}\right)_n}{\partial\boldsymbol{\xi}_{\text{c}}}\right]^H\right\},
\end{equation}
where $\overline{\boldsymbol{y}}$ represents the noise-less and interference-less part of the received signals at the MS.

\noindent\textbf{\emph{\underline{Lemma }1: }}\emph{The partial derivative term in $\boldsymbol{F}_{\text{c}}$ can be calculated as}
\begin{equation}
	\label{eq12}
	\frac{\partial\left(\overline{\boldsymbol{y}}\right)_n}{\partial\theta_{\text{D}}}=h_{\text{D}}\left(\frac{\partial\boldsymbol{b}\left(\theta_{\text{D}}\right)}{\partial\theta_{\text{D}}}\boldsymbol{a}^T\left(\theta_{\text{D}}^{\prime}\right)\boldsymbol{w}\right)_n,
\end{equation}
\begin{equation}
	\label{eq13}
	\frac{\partial\left(\overline{\boldsymbol{y}}\right)_n}{\partial\theta_{\text{R}}}=h_{\text{R}}\left(\frac{\partial\boldsymbol{c}(\theta_{\text{R}})\boldsymbol{b}^T(\theta_{\text{R}})}{\partial\theta_{\text{R}}}\boldsymbol{\Phi}\boldsymbol{b}\left(\theta_{\text{R}}^{\prime}\right)\boldsymbol{a}^T\left(\theta_{\text{R}}^{\prime}\right)\boldsymbol{w}\right)_n,
\end{equation}
\begin{equation}
	\label{eq14}
	\frac{\partial\left(\overline{\boldsymbol{y}}\right)_n}{\partial\Im\left\{h_{\text{D}}\right\}}=\left(j\boldsymbol{b}(\theta_{\text{D}})\boldsymbol{a}^T\left(\theta_{\text{D}}^{\prime}\right)\boldsymbol{w}\right)_n,
\end{equation}
\begin{equation}
	\label{eq15}
	\frac{\partial\left(\overline{\boldsymbol{y}}\right)_n}{\partial\Re\left\{h_{\text{D}}\right\}}=\left(\boldsymbol{b}(\theta_{\text{D}})\boldsymbol{a}^T\left(\theta_{\text{D}}^{\prime}\right)\boldsymbol{w}\right)_n,
\end{equation}
\begin{equation}
	\label{eq16}
	\frac{\partial\left(\overline{\boldsymbol{y}}\right)_n}{\partial\Im\left\{h_{\text{R}}\right\}}=\left(j\boldsymbol{b}(\theta_{\text{R}})\boldsymbol{c}^T(\theta_{\text{R}})\boldsymbol{\Phi}\boldsymbol{c}\left(\theta_{\text{R}}^{\prime}\right)\boldsymbol{a}^T\left(\theta_{\text{R}}^{\prime}\right)\boldsymbol{w}\right)_n,
\end{equation}
\begin{equation}
	\label{eq17}
	\frac{\partial\left(\overline{\boldsymbol{y}}\right)_n}{\partial\Re\left\{h_{\text{R}}\right\}}=\left(\boldsymbol{b}(\theta_{\text{R}})\boldsymbol{c}^T(\theta_{\text{R}})\boldsymbol{\Phi}\boldsymbol{c}\left(\theta_{\text{R}}^{\prime}\right)\boldsymbol{a}^T\left(\theta_{\text{R}}^{\prime}\right)\boldsymbol{w}\right)_n.
\end{equation}
\emph{Moreover, the Jacobian matrix $\boldsymbol{\Gamma}$ can be calculated as}
\begin{equation}
	\label{eq18}
	\boldsymbol{\Gamma}=\begin{bmatrix}
		\begin{matrix}
			\frac{-\vert y_{\text{SBS}}-y\vert}{\left(x-x_{\text{SBS}}\right)^2+\left(y-y_{\text{SBS}}\right)^2},& \frac{\vert x-x_{\text{SBS}}\vert}{\left(x-x_{\text{SBS}}\right)^2+\left(y-y_{\text{SBS}}\right)^2},\\
			\frac{-\vert y_{\text{RIS}}-y\vert}{\left(x-x_{\text{RIS}}\right)^2+\left(y-y_{\text{RIS}}\right)^2},& \frac{\vert x-x_{\text{RIS}}\vert}{\left(x-x_{\text{RIS}}\right)^2+\left(y-y_{\text{RIS}}\right)^2}, 
		\end{matrix}&\boldsymbol{0}_{2\times4},\\
		 \boldsymbol{0}_{2\times4},&\boldsymbol{\text{I}}_{4\times4}
	\end{bmatrix},
\end{equation}
\emph{where $\left(x_{\text{SBS}},y_{\text{SBS}}\right)$ and $\left(x_{\text{RIS}},y_{\text{RIS}}\right)$ denote the 2D positions of the SBS and the RIS, respectively.}\\
\emph{Proof:} Please refer to Appendix A.$\hfill\blacksquare$

\noindent\textbf{\emph{\underline{Remark }1: }}\emph{The SBS will gather the spectrum measurement information from all MSs to construct a REM, which can be used to alleviate the hidden node problem in CR networks. The REM is generated through interpolation methods, such as near neighbour (NN) and Kriging methods \cite{yilmaz2013radio}, where the spatial correlation of the positions of MSs is utilized. Therefore, the accuracy of the REM mainly depends on the accuracy of the positions of MSs. The accuracy of the REM can be improved by minimizing the PEBs of MSs.\\\indent As far as authors know, there is no effective joint beamforming method to minimize the PEB directly under some constraints, such as the constraints for the communication rate and the RIS unit-modulus. We use the SINR of the received signal at MS to characterize the localization performance of MS, as the PEB in (\ref{eq5}) and (\ref{eq9})-(\ref{eq18}) is inversely proportional to the SNR of the received signal.}

The average SINR of the $m$-th MS can be expressed as
\begin{equation}
	\label{eq19}
	\begin{aligned}
	\text{SINR}_{\text{MS},m}=\left(1-\frac{\tau}{T}\right)\Pr\left\{\mathcal{H}_0\right\}\left(1-P_f(\tau)\right)\text{SINR}_{\text{Case1},m}\\
	+\left(1-\frac{\tau}{T}\right)\Pr\left\{\mathcal{H}_1\right\}\left(1-P_d(\tau,\boldsymbol{\Phi})\right)\text{SINR}_{\text{Case2},m},
	\end{aligned}
\end{equation}
where Case 1 and Case 2 denote the two cases introduced in the last subsection. The SINR in the two cases can be expressed as
\begin{equation}
	\label{eq20}
	\text{SINR}_{\text{Case1},m}=
	\frac{\left\Vert\left(\boldsymbol{H}_{\text{R2MS},m}^T\boldsymbol{\Phi}\boldsymbol{H}_{\text{S2R}}^T+\boldsymbol{H}_{\text{S2MS},m}^T\right)\boldsymbol{w}_m\right\Vert_2^2}{\zeta_m+\sigma_{\text{MS}}^2}
\end{equation}
and
\begin{equation}
	\label{eq21}
	\begin{aligned}
	\text{SINR}_{\text{Case2},m}= \frac{\left\Vert\left(\boldsymbol{H}_{\text{R2MS},m}^T\boldsymbol{\Phi}\boldsymbol{H}_{\text{S2R}}^T+\boldsymbol{H}_{\text{S2MS},m}^T\right)\boldsymbol{w}_m\right\Vert_2^2}{\zeta_m+P_{\text{PBS}}\left\Vert\left(\boldsymbol{h}_{\text{P2MS},m}+\boldsymbol{H}_{\text{R2MS},m}^T\boldsymbol{\Phi}\boldsymbol{h}_{\text{P2R}}\right)\right\Vert_2^2+\sigma_{\text{MS}}^2}
	\end{aligned}
\end{equation}
where $\zeta_m=\sum\limits_{i=M+1}^{M+K}\left\Vert\left(\boldsymbol{H}_{\text{R2MS},m}^T\boldsymbol{\Phi}\boldsymbol{H}_{\text{S2R}}^T+\boldsymbol{H}_{\text{S2MS},m}^T\right)\boldsymbol{w}_i\right\Vert_2^2$ denotes the interference power from the SUEs. It is noted that the interference signals from the other MSs is eliminated due to the orthogonality of the sensing waveforms.

\subsection{Average Communication Rate}
The received signals of the $k$-th SUE can be written as
\begin{equation}
	\label{eq22}
	\begin{aligned}
	&y_{\text{SU},k}(t)=\left(\boldsymbol{h}_{\text{S2SU},k}^T+\boldsymbol{h}_{\text{R2SU},k}^T\boldsymbol{\Phi}\boldsymbol{H}_{\text{S2R}}^T\right)\boldsymbol{w}_{M+k}x_{M+k}(t)\\
	&+\sum_{i=1,i\neq M+k}^{M+K}\left(\boldsymbol{h}_{\text{S2SU},k}^T+\boldsymbol{h}_{\text{R2SU},k}^T\boldsymbol{\Phi}\boldsymbol{H}_{\text{S2R}}^T\right)\boldsymbol{w}_ix_{i}(t)\\
	&+\underbrace{\sqrt{P_{\text{PBS}}}\left(h_{\text{P2SU},k}+\boldsymbol{h}_{\text{R2SU},k}^T\boldsymbol{\Phi}\boldsymbol{h}_{\text{P2R}}\right)x_{\text{PBS}}(t)}_{\text{Case 2}}+n_{\text{SU}}(t),
	\end{aligned}
\end{equation}
where $n_{\text{SU}}(t)$ is the AWGN of the SUE with the variance $\sigma_{\text{SU}}^2$ and the third term on the right-hand side only exists when Case 2 introduced in Section \uppercase\expandafter{\romannumeral2}-A occurs. Then, the average communication rate of the $k$-th user can be expressed as
\begin{equation}
	\label{eq23}
	\begin{aligned}
		&R_k=\left(1-\frac{\tau}{T}\right)\Pr\left\{\mathcal{H}_0\right\}\left(1-P_f(\tau)\right)\log_2\left(1+\text{SINR}_{\text{Case1},k}^{\prime}\right)\\
		&+\left(1-\frac{\tau}{T}\right)\Pr\left\{\mathcal{H}_1\right\}\left(1-P_d(\tau,\boldsymbol{\Phi})\right)\log_2\left(1+\text{SINR}_{\text{Case2},k}^{\prime}\right),
	\end{aligned}
\end{equation}
where the SINRs in the two cases can be expressed as
\begin{equation}
	\label{eq24}
	\text{SINR}_{\text{Case1},k}^{\prime}=
	\frac{\left\vert\left(\boldsymbol{h}_{\text{S2SU},k}^T+\boldsymbol{h}_{\text{R2SU},k}^T\boldsymbol{\Phi}\boldsymbol{H}_{\text{S2R}}^T\right)\boldsymbol{w}_{M+k}\right\vert^2}{\zeta_k^{\prime}+\sigma_{\text{SU}}^2}
\end{equation}
and
\begin{equation}
	\label{eq25}
	\text{SINR}_{\text{Case2},k}^{\prime}=\frac{\left\vert\left(\boldsymbol{h}_{\text{S2SU},k}^T+\boldsymbol{h}_{\text{R2SU},k}^T\boldsymbol{\Phi}\boldsymbol{H}_{\text{S2R}}^T\right)\boldsymbol{w}_{M+k}\right\vert^2}{\zeta_k^{\prime}+P_{\text{PBS}}\left\vert h_{\text{P2SU},k}+\boldsymbol{h}_{\text{R2SU},k}^T\boldsymbol{\Phi}\boldsymbol{h}_{\text{P2R}}\right\vert^2+\sigma_{\text{SU}}^2},
\end{equation}
where $\zeta_k^{\prime}=\sum\limits_{i=M+1,i\neq k}^{M+K}\left\vert\left(\boldsymbol{h}_{\text{S2SU},k}^T+\boldsymbol{h}_{\text{R2SU},k}^T\boldsymbol{\Phi}\boldsymbol{H}_{\text{S2R}}^T\right)\boldsymbol{w}_i\right\vert^2$ is the interference power from the other SUEs. We assume that the SUEs have prior information on the sensing waveforms of the MSs since the SUEs and the MSs both belong to the secondary system. Therefore, the SUEs can also eliminate the interference signals from the MSs. Additionally, the interference power of the $l$-th PUE can be expressed as
\begin{equation}
	\label{eq26}
	\gamma_l=\sum\limits_{i=1}^{M+K}\left\vert\left(\boldsymbol{h}_{\text{S2PU},l}^T+\boldsymbol{h}_{\text{R2PU},l}^T\boldsymbol{\Phi}\boldsymbol{H}_{\text{S2R}}^T\right)\boldsymbol{w}_i\right\vert^2,
\end{equation}
where we assume that the PUEs have no prior information on the sensing waveforms of the MSs.

\subsection{Problem Formulation}
In this subsection, we formulate the below max-min problem, i.e., (P1), to minimize the weighted PEBs of MSs while satisfing the performance of SUEs and PUEs by designing the SS time, the BS beamforming matrix, and the RIS beamforming matrix.
\begin{subequations}
	\label{eq27}
	\begin{align}
		&\text{(P1)}\mathop{\text{max min}}\limits_{\tau,\boldsymbol{W},\boldsymbol{\Phi}}\  \lambda_m\text{SINR}_{\text{MS},m}\left(\tau,\boldsymbol{W},\boldsymbol{\Phi}\right) \label{eq27a} \tag{27a} \\
		&\qquad\text{s.t.}\  R_{k}\left(\tau,\boldsymbol{W},\boldsymbol{\Phi}\right) \geq r_{k}, \forall k \in \mathcal{K}, \label{eq27b} \tag{27b} \\
		&\qquad\quad\  P_d\left(\tau,\boldsymbol{\Phi}\right)\geq P_{d,0}, \label{eq27c} \tag{27c} \\
		&\qquad\quad\  P_f(\tau)\leq P_{f,0}, \label{eq27d} \tag{27d}\\
		&\qquad\quad \left(1-\frac{\tau}{T}\right)\big[\Pr\left\{\mathcal{H}_0\right\}\left(1-P_f(\tau)\right)+\Pr\left\{\mathcal{H}_1\right\}\nonumber\\
		&\qquad\quad\ \cdot\left(1-P_d(\tau,\boldsymbol{\Phi})\right)\big]\sum_{i=1}^{M+K}\text{Tr}\left(\boldsymbol{w}_i\boldsymbol{w}_i^H\right)\leq P_{\text{SBS}}, \label{eq27e} \tag{27e}\\
		&\qquad\quad\left(1-\frac{\tau}{T}\right)\Pr\left\{\mathcal{H}_1\right\}\left(1-P_d(\tau,\boldsymbol{\Phi})\right)\gamma_l\leq\gamma_{l,0},\forall l\in\mathcal{L}, \label{eq27f} \tag{27f}\\
		&\qquad\quad\ \left\vert\boldsymbol{\Phi}\right\vert_{n,n}=1,\forall n\in\mathcal{N}, \label{eq27g} \tag{27g}
	\end{align}
\end{subequations}
where $\lambda_m$ in (27a) is the weighted factor of the $m$-th MS; the constraints in (\ref{eq27b}) ensure the communication rate of each SUE being greater than the threshold $r_k$; the constraints in (\ref{eq27c}) and (\ref{eq27d}) ensure the detection probability being greater than the threshold $P_{d,0}$ and the probability of false alarm being smaller than the threshold $P_{f,0}$; the constraint in (\ref{eq27e}) is the transmit power budget of the SBS; the constraints in (\ref{eq27f}) ensure the interference power at each PUE being smaller than the threshold $\gamma_{l,0}$; and the constraints in (\ref{eq27g}) is the unit-modulus constraints of the RIS. (P1) is a highly non-convex problem due to the fact that its objective function and constraints are all non-convex and the variables $\tau,\boldsymbol{W},\boldsymbol{\Phi}$ are coupled together. Therefore, a BCD-based algorithm is proposed in the next section to solve (P1).

\section{Algorithm Design}
A BCD-based algorithm is proposed in this section to decouple the variables and solve the non-convex (P1). Specifically, the Dinkelbach's transform and the SCA methods are used to solve the decoupled sub-problems.

\subsection{Spectrum Sensing Time Design}
In this subsection, we fix the transmit beamforming and the RIS beamforming matrices to design the SS time. We use an exhaustive search method to obtain the optimal SS time since the SS time is a one-dimensional variable. The exhaustive search method is performed as the SS time exhaustively takes a value from 0 to $T$. During the exhaustion, we first evaluate whether the constraints in (\ref{eq27b})-(\ref{eq27f}) are satisfied and then calculate the objective function in (\ref{eq27a}). The optimal SS time is the value that maximizes the sensing SINR and satisfies all the constraints.

\subsection{Transmit Beamforming Design}
To design the transmit beamforming matrix, the SS time and the RIS beamforming matrix are kept fixed in this subsection. The max-min (P1) can be reformulated as the following semidefinite program (SDP) problem.
\begin{subequations}
	\label{eq28}
	\begin{align}
		\text{(P2)}\mathop{\text{max}}\limits_{t,\left\{\boldsymbol{\overline{W}}_i\right\}}\ & t \label{eq28a} \tag{28a} \\
		\text{s.t.} \quad& \frac{a_{1,m}\text{Tr}\left(\boldsymbol{G}_m\boldsymbol{\overline{W}}_m\boldsymbol{G}_m^H\right)}{\overline{\zeta}_m+\sigma_{\text{MS}}^2}\nonumber\\
		&\!\!\!+\frac{a_{2,m}\text{Tr}\left(\boldsymbol{G}_m\boldsymbol{\overline{W}}_m\boldsymbol{G}_m^H\right)}{\overline{\zeta}_m+\overline{a}_m}\geq t,\forall m\in\mathcal{M}, \label{eq28b} \tag{28b} \\
		&b\log_2\left(1+\frac{\text{Tr}\left(\boldsymbol{g}_k\boldsymbol{g}_k^H\boldsymbol{\overline{W}}_{M+k}\right)}{\overline{\zeta}_k^{\prime}+\sigma_{\text{SU}}^2}\right)+\widetilde{b}\nonumber \\
		&\!\!\!\!\!\cdot\log_2\left(1+\frac{\text{Tr}\left(\boldsymbol{g}_k\boldsymbol{g}_k^H\boldsymbol{\overline{W}}_{M+k}\right)}{\overline{\zeta}_k^{\prime}+\widetilde{a}_k}\right)\geq r_k,\forall k\in\mathcal{K},\label{eq28c} \tag{28c}\\
		&\left(b+\widetilde{b}\right)\sum\limits_{i=1}^{M+K}\text{Tr}\left(\overline{\boldsymbol{W}}_i\right)\leq P_{\text{SBS}}, \label{eq28d} \tag{28d}\\
		&\widetilde{b}\sum\limits_{i=1}^{M+K}\text{Tr}\left(\overline{\boldsymbol{g}}_l\overline{\boldsymbol{g}}_l^H\overline{\boldsymbol{W}}_i\right)\leq \gamma_{l,0},\forall l\in\mathcal{L}, \label{eq28e} \tag{28e}\\
		&\text{rank}\left(\overline{\boldsymbol{W}}_i\right)=1,\forall i\in\mathcal{M}\cup\mathcal{K}, \label{eq28f} \tag{28f}\\
		&\overline{\boldsymbol{W}}_i\succeq0,\forall i\in\mathcal{M}\cup\mathcal{K}, \label{eq28g} \tag{28g}
	\end{align}
\end{subequations}
where $\boldsymbol{\overline{W}}_i=\boldsymbol{w}_i\boldsymbol{w}_i^H$, $b=\left(1-\frac{\tau}{T}\right)\Pr\left\{\mathcal{H}_0\right\}\left(1-P_f\right)$, $\widetilde{b}=\left(1-\frac{\tau}{T}\right)\Pr\left\{\mathcal{H}_1\right\}\left(1-P_d\right)$, $a_{1,m}=\lambda_mb$, and $a_{2,m}=\lambda_m\widetilde{b}$. Moreover,
\begin{equation}
	\label{eq29}
	\boldsymbol{G}_m=\boldsymbol{H}_{\text{R2MS},m}^T\boldsymbol{\Phi}\boldsymbol{H}_{\text{S2R}}^T+\boldsymbol{H}_{\text{S2MS},m},
\end{equation} 
\begin{equation}
	\label{eq30}
	\boldsymbol{g}_k=\boldsymbol{h}_{\text{S2SU},k}^*+\boldsymbol{H}_{\text{S2R}}^*\boldsymbol{\Phi}\boldsymbol{h}_{\text{R2SU},k}^*,
\end{equation}
\begin{equation}
	\label{eq31}
	\overline{\boldsymbol{g}}_l=\boldsymbol{h}_{\text{S2PU},l}^*+\boldsymbol{H}_{\text{S2R}}^*\boldsymbol{\Phi}\boldsymbol{h}_{\text{R2PU},l}^*,
\end{equation}
\begin{equation}
	\label{eq32}
	\overline{a}_m=P_{\text{PBS}}\left\Vert\left(\boldsymbol{h}_{\text{P2MS},m}+\boldsymbol{H}_{\text{R2MS},m}^T\boldsymbol{\Phi}\boldsymbol{h}_{\text{P2R}}\right)\right\Vert_2^2+\sigma_{\text{MS}}^2,
\end{equation}
\begin{equation}
	\label{eq33}
	\widetilde{a}_k=P_{\text{PBS}}\left\vert h_{\text{P2SU},k}+\boldsymbol{h}_{\text{R2SU},k}^T\boldsymbol{\Phi}\boldsymbol{h}_{\text{P2R}}\right\vert^2+\sigma_{\text{SU}}^2,
\end{equation}
\begin{equation}
	\label{eq34}
	\overline{\zeta}_m=\sum_{i=M+1}^{M+K}\text{Tr}\left(\boldsymbol{G}_m\boldsymbol{\overline{W}}_i\boldsymbol{G}_m^H\right),
\end{equation}
and
\begin{equation}
	\label{eq35}
	\overline{\zeta}_k^{\prime}=\sum_{i=1}^K\text{Tr}\left(\boldsymbol{g}_k\boldsymbol{g}_k^H\boldsymbol{\overline{W}}_{M+i}\right).
\end{equation}
(P2) is non-convex due to the non-convex constraints in (\ref{eq28b}), (\ref{eq28c}), and (\ref{eq28f}). We first use Dinkelbach’s transform \cite{schaible1976fractional} to transform the constraints of the sensing SINR, i.e., the constraints in (\ref{eq28b}), to convex constraints. The constraint of the sensing SINR for the $m$-th MS can be transformed as
\begin{equation}
	\label{eq36}
	\begin{aligned}
	a_{1,m}\text{Tr}\left(\boldsymbol{G}_m\overline{\boldsymbol{W}}_m\boldsymbol{G}_m^H\right)-y_{1,m}\left(\overline{\zeta}_m+\sigma_{\text{MS}}^2\right)+ a_{2,m}\text{Tr}\left(\boldsymbol{G}_m\overline{\boldsymbol{W}}_m\boldsymbol{G}_m^H\right)-y_{2,m}\left(\overline{\zeta}_m+\overline{a}_m\right)\geq t,		
	\end{aligned}
\end{equation}
where $y_{1,m}$ and $y_{2,m}$ are the auxiliary variables. We can iteratively update the transmit beamforming matrix and the auxiliary variables. The auxiliary variables are updated by
\begin{equation}
	\label{eq37}
	\begin{aligned}
		&y_{1,m}[i+1]=\frac{a_{1,m}\text{Tr}\left(\boldsymbol{G}_m\overline{\boldsymbol{W}}_m[i]\boldsymbol{G}_m^H\right)}{\overline{\zeta}_m[i]+\sigma_{\text{MS}}^2},\\
		&y_{2,m}[i+1]=\frac{a_{2,m}\text{Tr}\left(\boldsymbol{G}_m\overline{\boldsymbol{W}}_m[i]\boldsymbol{G}_m^H\right)}{\overline{\zeta}_m[i]+\overline{a}_m},
	\end{aligned}
\end{equation}
where $i$ denotes the number of Dinkelbach’s transform iteration. The constraints in (\ref{eq28b}) are convex when we keep $y_{1,m}$ and $y_{2,m}$ fixed.

\noindent\textbf{\emph{\underline{Theorem }1: }}\emph{The problem after Dinkelbach’s transform is equivalent to (P2) and is guaranteed to convergent.}\\
\emph{Proof:} We assume the optimal beamforming matrices of the problem after Dinkelbach’s transform are not those of (P2), then there exist beamforming matrices that can obtain higher sensing SINRs. This contradicts the assumption. Hence, the two problems are equivalent. Moreover, the auxiliary variables $y_{1,m}$ and $y_{2,m}$ are nondecreasing after each iteration. Therefore, the convergence of the problem after Dinkelbach’s transform is guaranteed. This completes the proof.$\hfill\blacksquare$

We then use the SCA method to deal with the constraints of the communication rates of SUEs, i.e., the constraints in (\ref{eq28c}). The constraint of the communication rate for the $k$-th SUE can be reformulated as
\begin{equation}
	\label{eq38}
	\begin{aligned}
		&b\log_2\left(\overline{\zeta}_k^{\prime}\!+\!\text{Tr}\left(\boldsymbol{g}_k\boldsymbol{g}_k^H\boldsymbol{\overline{W}}_{M+k}\right)\!+\!\sigma_{\text{SU}}^2\right)\!-\!b\log_2\left(\overline{\zeta}_k^{\prime}\!+\!\sigma_{\text{SU}}^2\right)\\
		&+\widetilde{b}\log_2\left(\overline{\zeta}_k^{\prime}\!+\!\text{Tr}\left(\boldsymbol{g}_k\boldsymbol{g}_k^H\boldsymbol{\overline{W}}_{M+k}\right)\!+\!\widetilde{a}_k\right)\!-\!\widetilde{b}\log_2\left(\overline{\zeta}_k^{\prime}\!+\!\widetilde{a}_k\right)\!\geq \!r_k.
	\end{aligned}
\end{equation}
The non-convex terms in (\ref{eq38}), i.e., the second and the fourth terms, can be relaxed using the first order Taylor expansion, which can be expressed as
\begin{equation}
	\label{eq39}
	\begin{aligned}
	b\log_2\left(\overline{\zeta}_k^{\prime}+\sigma_{\text{SU}}^2\right)&\leq b\log_2\left(\overline{\zeta}_k^{\prime}[j-1]+\sigma_{\text{SU}}^2\right)\\
	&+\frac{b\sum_{i=1,i\neq k}^{K}\text{Tr}\left[\boldsymbol{g}_k\boldsymbol{g}_k^H\left(\overline{\boldsymbol{W}}_{M+i}-\overline{\boldsymbol{W}}_{M+i}[j-1]\right)\right]}{\ln2\left[\sum_{i=1,i\neq k}^{K}\text{Tr}\left(\boldsymbol{g}_k\boldsymbol{g}_k^H\overline{\boldsymbol{W}}_{M+i}[j-1]\right)+\sigma_{\text{SU}}^2\right]}
	\end{aligned}
\end{equation}
and
\begin{equation}
	\label{eq40}
	\begin{aligned}
	\widetilde{b}\log_2\left(\overline{\zeta}_k^{\prime}+\widetilde{a}_k\right)&\leq \widetilde{b}\log_2\left(\overline{\zeta}_k^{\prime}[j-1]+\widetilde{a}_k\right)\\
	&+\frac{\widetilde{b}\sum_{i=1,i\neq k}^{K}\text{Tr}\left[\boldsymbol{g}_k\boldsymbol{g}_k^H\left(\overline{\boldsymbol{W}}_{M+i}-\overline{\boldsymbol{W}}_{M+i}[j-1]\right)\right]}{\ln2\left[\sum_{i=1,i\neq k}^{K}\text{Tr}\left(\boldsymbol{g}_k\boldsymbol{g}_k^H\overline{\boldsymbol{W}}_{M+i}[j-1]\right)+\widetilde{a}_k\right]},
	\end{aligned}
\end{equation}
where $j$ denotes the number of SCA iteration.

Furthermore, the semidefinite relaxation (SDR) method is used to relax the rank one constraints in (\ref{eq28f}). (P2) can be reformulated as
\begin{subequations}
	\label{eq41}
	\begin{align}
		\text{(P3)}\mathop{\text{max}}\limits_{t,\left\{\boldsymbol{\overline{W}}_i\right\}}\ & t \label{eq41a} \tag{41a} \\
		\text{s.t.} \quad& \text{(\ref{eq36}), (\ref{eq38})-(\ref{eq40}), (\ref{eq28d}), (\ref{eq28e}), (\ref{eq28g})}.\label{eq41b} \tag{41b} 
	\end{align}
\end{subequations}
Now, (P2) is a standard SDP problem, which can be solve by the CVX tool. However, the SDR method can not guarantee to obtain rank one beamforming matrices. According to Theorem 1 in \cite{xu2020resource}, the tightness of the SDR method applying to (P3) can be proved through Karush-Kuhn-Tucker (KKT) conditions. The scheme to solve the transmit beamforming problem is presented in \textbf{Algorithm 1}. 

\subsection{RIS Beamforming Design}
To design the RIS beamforming matrix, we keep the SS time and the transmit beamforming matrix fixed in this subsection. (P1) can be reformulated as the following SDP problem.
\begin{subequations}
	\label{eq42}
	\begin{align}
		\text{(P4)}&\mathop{\text{max}}\limits_{t,\boldsymbol{\bar{\Phi}}}\ t \label{eq42a} \tag{42a} \\
		\text{s.t.}& \  \frac{a_{1,m}\text{Tr}\left(\boldsymbol{\overline{G}}_{m,m}\boldsymbol{\bar{\Phi}}\right)}{\sum\limits_{i=M+1}^{M+K}\text{Tr}\left(\boldsymbol{\overline{G}}_{m,i}\boldsymbol{\bar{\Phi}}\right)+\sigma_{\text{MS}}^2}\nonumber\\
		&\!+\frac{c_m\left(1-P_d\left(\boldsymbol{\bar{\Phi}}\right)\right)\text{Tr}\left(\boldsymbol{\overline{G}}_{m,m}\boldsymbol{\bar{\Phi}}\right)}{\sum\limits_{i=M+1}^{M+K}\text{Tr}\left(\boldsymbol{\overline{G}}_{m,i}\boldsymbol{\bar{\Phi}}\right)+d_m}\geq t,\forall m\in\mathcal{M}, \label{eq42b} \tag{42b} \\
		&b\log_2\left(1+\frac{\text{Tr}\left(\boldsymbol{E}_{k,k}\boldsymbol{\bar{\Phi}}^*\right)}{\sum_{i=1,i\neq k}^{K}\text{Tr}\left(\boldsymbol{E}_{k,i}\boldsymbol{\bar{\Phi}}^*\right)+\sigma_{\text{SU}}^2}\right)+\widetilde{c}\nonumber \\
		&\!\!\!\!\!\!\cdot\left(1-P_d\left(\boldsymbol{\bar{\Phi}}\right)\right)\log_2\left(1+\frac{\text{Tr}\left(\boldsymbol{E}_{k,k}\boldsymbol{\bar{\Phi}}^*\right)}{\sum_{i=1,i\neq k}^{K}\text{Tr}\left(\boldsymbol{E}_{k,i}\boldsymbol{\bar{\Phi}}^*\right)+\overline{d}_k}\right)\nonumber\\
		&\geq r_k,\forall k\in\mathcal{K},\label{eq42c} \tag{42c}\\
		&P_d\left(\boldsymbol{\bar{\Phi}}\right)\geq P_{d,0},\label{eq42d} \tag{42d}\\
		&\tilde{c}\left(1-P_d\left(\boldsymbol{\bar{\Phi}}\right)\right)\sum_{i=1}^{M+K}\text{Tr}\left(\boldsymbol{\overline{E}}_{l,i}\boldsymbol{\bar{\Phi}}\right)\leq \gamma_{l,0},\forall l\in\mathcal{L},\label{eq42e} \tag{42e}\\
		&\text{rank}\left(\boldsymbol{\bar{\Phi}}\right)=1,\label{eq42f} \tag{42f}\\
		&\left\vert\boldsymbol{\bar{\Phi}}\right\vert_{n,n}=1,n\in\left\{1,\cdots,N_{\text{R}}+1\right\},\label{eq42g} \tag{42g}\\
		&\boldsymbol{\bar{\Phi}}\succeq 0,\label{eq42h} \tag{42h}		
	\end{align}
\end{subequations}

\begin{algorithm}[t]
	\caption{Dinkelbach’s transform and SCA methods for solving (P2).}\label{alg:alg1}
	\small
	\begin{algorithmic}
		\STATE 
		\STATE 1. $ \textbf{Inputs: }$$\tau$, $\boldsymbol{\Phi}$, $N_\text{B}$, $N_\text{R}$, $N_\text{M}$, $M$, $K$, $L$, $\Pr\left\{\mathcal{H}_0\right\}$, $\Pr\left\{\mathcal{H}_1\right\}$, $T$, $r_k$, $P_{\text{SBS}}$, $P_{\text{PBS}}$, $\gamma_{l,0}$, $\lambda_m$, $\sigma_{\text{MS}}^2$, $\sigma_{\text{SU}}^2$, $\sigma_{\text{SBS}}^2$, $i_{\text{max}}$, $j_{\text{max}}$, $\varepsilon$, all channel matrices and vectors. 
		\STATE 2. $ \textbf{Outputs: }$Beamforming matrix $\boldsymbol{W}$ for (P2).
		\STATE 3. $ \textbf{Initialization: }$Set $t=0$, solve (P2) to obtain $\boldsymbol{W}_{\text{DT}}[0]$, which represent the initial transmit beamforming matrix of Dinkelbach's transform, and calculate the sensing SINR as $O_{\text{DT}}[0]$. Set $O_{\text{DT}}[1]=+\inf$. Set $i,j=1$.
		\WHILE {$i\leq i_{\text{max}}$ and $\frac{\vert O_{\text{DT}}[i]-O_{\text{DT}}[i-1]\vert}{\vert O_{\text{DT}}[i-1]\vert}\geq\varepsilon$}
		\STATE 4.Calculate $y_{1,m}[i]$ and $y_{2,m}[i]$ according to (\ref{eq37}).
		\STATE 5.Set $\boldsymbol{W}_{\text{SCA}}[0]=\boldsymbol{W}_{\text{DT}}[i-1]$ and $O_{\text{SCA}}[1]=+\inf$.
		\WHILE {$j\leq j_{\text{max}}$ and $\frac{\vert O_{\text{SCA}}[j]-O_{\text{SCA}}[j-1]\vert}{\vert O_{\text{SCA}}[j-1]\vert}\geq\varepsilon$}
		\STATE 6.Solve (P3) to obtain $\boldsymbol{W}_{\text{SCA}}[j]$ and $O_{\text{SCA}}[j]$.
		\ENDWHILE
		\STATE 6.Set $\boldsymbol{W}_{\text{DT}}[i]=\boldsymbol{W}_{\text{SCA}}[j]$ and $O_{\text{DT}}[i]=O_{\text{SCA}}[j]$.
		\ENDWHILE
		\STATE 7.Set $\boldsymbol{W}=\boldsymbol{W}_{\text{DT}}[i]$.
	\end{algorithmic}
	\label{alg1}
\end{algorithm}
\noindent where $\boldsymbol{\bar{\Phi}}=\boldsymbol{\bar{\phi}}\boldsymbol{\bar{\phi}}^H$ with $\boldsymbol{\bar{\phi}}=\left(\boldsymbol{\phi}^T,1\right)^T$ and $\boldsymbol{\phi}=\text{diag}\left(\boldsymbol{\Phi}\right)$, $c_m=\lambda_m\left(1-\frac{\tau}{T}\right)\Pr\left\{\mathcal{H}_1\right\}$, $\widetilde{c}=\left(1-\frac{\tau}{T}\right)\Pr\left\{\mathcal{H}_1\right\}$,
\begin{equation}
	\label{eq43}
	d_m=P_{\text{PBS}}\text{Tr}\left(\boldsymbol{\widetilde{G}}_{m}\boldsymbol{\bar{\Phi}}\right)+\sigma_{\text{MS}}^2,
\end{equation}
\begin{equation}
	\label{eq44}
	\begin{aligned}
			\boldsymbol{\overline{G}}_{m,i}=\begin{bmatrix}
			\text{diag}\left(\boldsymbol{w}_i^H\boldsymbol{H}_{\text{S2R}}^*\right)\boldsymbol{H}_{\text{R2MS},m}^*\boldsymbol{H}_{\text{R2MS},m}^T\text{diag}\left(\boldsymbol{H}_{\text{S2R}}^T\boldsymbol{w}_i\right),\\
			\text{diag}\left(\boldsymbol{w}_i^H\boldsymbol{H}_{\text{S2R}}^*\right)\boldsymbol{H}_{\text{R2MS},m}^*\boldsymbol{H}_{\text{S2MS},m}^T\boldsymbol{w}_{i};\\
			\boldsymbol{w}_i^H\boldsymbol{H}_{\text{S2MS},m}^*\boldsymbol{H}_{\text{R2MS},m}^T\text{diag}\left(\boldsymbol{H}_{\text{S2R}}^T\boldsymbol{w}_i\right),\\
			\boldsymbol{w}_i\boldsymbol{H}_{\text{S2MS},m}^*\boldsymbol{H}_{\text{S2MS},m}^T\boldsymbol{w}_i
		\end{bmatrix},
	\end{aligned}
\end{equation}
\begin{equation}
	\label{eq45}
	\begin{aligned}
	&\boldsymbol{\widetilde{G}}_{m}=\begin{bmatrix}
		\text{diag}\left(\boldsymbol{h}_{\text{P2R}}^H\right)\boldsymbol{H}_{\text{R2MS},m}^*\boldsymbol{H}_{\text{R2MS},m}^T\text{diag}\left(\boldsymbol{h}_{\text{P2R}}\right),\\
		\text{diag}\left(\boldsymbol{h}_{\text{P2R}}^H\right)\boldsymbol{H}_{\text{R2MS},m}^*\boldsymbol{h}_{\text{P2MS},m};\\
		\boldsymbol{h}_{\text{P2MS},m}^H\boldsymbol{H}_{\text{R2MS},m}^T\text{diag}\left(\boldsymbol{h}_{\text{P2R}}\right),\\
		\boldsymbol{h}_{\text{P2MS},m}^H\boldsymbol{h}_{\text{P2MS},m}
	\end{bmatrix},
	\end{aligned}
\end{equation}
\begin{equation}
	\label{eq46}
	\overline{d}_k=P_{\text{PBS}}\text{Tr}\left(\boldsymbol{\overline{E}}_{k}\boldsymbol{\bar{\Phi}}\right)+\sigma_{\text{SU}}^2,
\end{equation}
\begin{equation}
	\label{eq47}
	\begin{aligned}
		&\boldsymbol{{E}}_{k,i}=\begin{bmatrix}
			\text{diag}\left(\boldsymbol{h}_{\text{R2SU},k}\right)\boldsymbol{H}_{\text{S2R}}^T\boldsymbol{w}_{M+i}\boldsymbol{w}_{M+i}^H\boldsymbol{H}_{\text{S2R}}^*\text{diag}\left(\boldsymbol{h}_{\text{R2SU},k}^H\right),\\
			\text{diag}\left(\boldsymbol{h}_{\text{R2SU},k}\right)\boldsymbol{H}_{\text{S2R}}^T\boldsymbol{w}_{M+i}\boldsymbol{w}_{M+i}^H\boldsymbol{h}_{\text{S2SU},k}^*;\\
			\boldsymbol{h}_{\text{S2SU},k}^T\boldsymbol{w}_{M+i}\boldsymbol{w}_{M+i}^H\boldsymbol{H}_{\text{S2R}}^*\text{diag}\left(\boldsymbol{h}_{\text{R2SU},k}^H\right),\\
			\boldsymbol{h}_{\text{S2SU},k}^T\boldsymbol{w}_{M+i}\boldsymbol{w}_{M+i}^H\boldsymbol{h}_{\text{S2SU},k}^*
		\end{bmatrix},
	\end{aligned}
\end{equation}
\begin{equation}
	\label{eq48}
	\begin{aligned}
	\boldsymbol{\overline{E}}_{k}=\begin{bmatrix}
		\text{diag}\left(\boldsymbol{h}_{\text{P2R}}^H\right)\boldsymbol{h}_{\text{R2SU},k}^*\boldsymbol{h}_{\text{R2SU},k}^T\text{diag}\left(\boldsymbol{h}_{\text{P2R}}\right),\\
		\text{diag}\left(\boldsymbol{h}_{\text{P2R}}^H\right)\boldsymbol{h}_{\text{R2SU},k}^*\boldsymbol{h}_{\text{P2SU},k};\\
		\boldsymbol{h}_{\text{P2SU},k}^*\boldsymbol{h}_{\text{R2SU},k}^T\text{diag}\left(\boldsymbol{h}_{\text{P2R}}\right),\\
		\boldsymbol{h}_{\text{P2SU},k}^*\boldsymbol{h}_{\text{P2SU},k}
	\end{bmatrix},
	\end{aligned}
\end{equation}
and
\begin{equation}
	\label{eq49}
	\begin{aligned}
		\boldsymbol{\widetilde{E}}_{l,i}=\begin{bmatrix}
			\text{diag}\left(\boldsymbol{w}_i^H\boldsymbol{H}_{\text{S2R}}^*\right)\boldsymbol{h}_{\text{R2PU},l}^*\boldsymbol{h}_{\text{R2PU},l}^T\text{diag}\left(\boldsymbol{H}_{\text{S2R}}^T\boldsymbol{w}_i\right),\\
			\text{diag}\left(\boldsymbol{w}_i^H\boldsymbol{H}_{\text{S2R}}^*\right)\boldsymbol{h}_{\text{R2PU},l}^*\boldsymbol{h}_{\text{S2PU},l}^T\boldsymbol{w}_i;\\
			\boldsymbol{w}_i^H\boldsymbol{h}_{\text{S2PU},l}^*\boldsymbol{h}_{\text{R2PU},l}^T\text{diag}\left(\boldsymbol{H}_{\text{S2R}}^T\boldsymbol{w}_i\right),\\
			\boldsymbol{w}_i^H\boldsymbol{h}_{\text{S2PU},l}^*\boldsymbol{h}_{\text{S2PU},l}^T\boldsymbol{w}_i
		\end{bmatrix}.
	\end{aligned}
\end{equation}
(P4) is non-convex due to the constraints in (\ref{eq42b})-(\ref{eq42f}) are non-convex. We first relax the detection probability using the following inequality \cite{chiani2003new}
\begin{equation}
	\label{eq50}
	\begin{aligned}
		P_d\left(\boldsymbol{\bar{\Phi}}\right)&\leq1-\frac{1}{12}\exp\left\{-\frac{1}{2}\left[\frac{\epsilon-\sigma_{\text{SBS}}^2\left(N_{\text{B}}+\text{SNR}_{\text{SBS}}\right)}{\left(\sigma_{\text{SBS}}^2/\sqrt{\tau f_s}\right)\sqrt{2\text{SNR}_{\text{SBS}}+N_{\text{B}}}}\right]^2\right\}\\
		&-\frac{1}{4}\exp\left\{-\frac{2}{3}\left[\frac{\epsilon-\sigma_{\text{SBS}}^2\left(N_{\text{B}}+\text{SNR}_{\text{SBS}}\right)}{\left(\sigma_{\text{SBS}}^2/\sqrt{\tau f_s}\right)\sqrt{2\text{SNR}_{\text{SBS}}+N_{\text{B}}}}\right]^2\right\}.
	\end{aligned}
\end{equation}
The Fourier series with an optimal sampling interval is used in \cite{chiani2003new} to derive the above inequality. The expression on the right side of (\ref{eq50}) is still non-convex. We introduce an auxiliary variable $z$ to deal with it. The detection probability can be transformed as
\begin{equation}
	\label{eq51}
	P_d\left(\boldsymbol{\bar{\Phi}}\right)=1-\frac{1}{12}\exp\left\{-\frac{z}{2}\right\}-\frac{1}{4}\exp\left\{-\frac{2z}{3}\right\},
\end{equation}
and $z$ is constrained by
\begin{equation}
	\label{eq52}
	z\leq \left[\frac{\epsilon-\sigma_{\text{SBS}}^2\left(N_{\text{B}}+P_{\text{PBS}}\text{Tr}\left(\hat{\boldsymbol{E}}\boldsymbol{\bar{\Phi}}\right)/\sigma_{\text{SBS}}^2\right)}{\left(\sigma_{\text{SBS}}^2/\sqrt{\tau f_s}\right)\sqrt{2P_{\text{PBS}}\text{Tr}\left(\hat{\boldsymbol{E}}\boldsymbol{\bar{\Phi}}\right)/\sigma_{\text{SBS}}^2+N_{\text{B}}}}\right]^2,
\end{equation}
where
\begin{equation}
	\label{eq53}
	\begin{aligned}
		\hat{\boldsymbol{E}}=\begin{bmatrix}
			\text{diag}(\boldsymbol{h}_{\text{P2R}}^H)\boldsymbol{H}_{\text{S2R}}^H\boldsymbol{H}_{\text{S2R}}\text{diag}(\boldsymbol{h}_{\text{P2R}}),&\!\!\!\text{diag}(\boldsymbol{h}_{\text{P2R}}^H)\boldsymbol{H}_{\text{S2R}}^H\boldsymbol{h}_{\text{P2S}}\\
			\boldsymbol{h}_{\text{P2S}}^H\boldsymbol{H}_{\text{S2R}}\text{diag}\left(\boldsymbol{h}_{\text{P2R}}\right),&\boldsymbol{h}_{\text{P2S}}^H\boldsymbol{h}_{\text{P2S}}			
		\end{bmatrix}.
	\end{aligned}
\end{equation}
Equation (\ref{eq52}) is rewritten as
\begin{equation}
	\label{eq54}
	\begin{aligned}
	\sigma_{\text{SBS}}^2\left(N_{\text{B}}+P_{\text{PBS}}\text{Tr}\left(\hat{\boldsymbol{E}}\boldsymbol{\bar{\Phi}}\right)/\sigma_{\text{SBS}}^2\right)-\sigma_{\text{SBS}}^2\sqrt{\frac{z}{\tau f_s}\left(2P_{\text{PBS}}\text{Tr}\left(\hat{\boldsymbol{E}}\boldsymbol{\bar{\Phi}}\right)/\sigma_{\text{SBS}}^2+N_{\text{B}}\right)}\geq \epsilon,
	\end{aligned}
\end{equation}
where the second term on the left side of the inequality is non-convex. The SCA method can approximate the second term as
\begin{equation}
	\label{eq55}
	\begin{aligned}
		\sigma_{\text{SBS}}^2\sqrt{\frac{z}{\tau f_s}\left(2P_{\text{PBS}}\text{Tr}\left(\hat{\boldsymbol{E}}\boldsymbol{\bar{\Phi}}\right)/\sigma_{\text{SBS}}^2+N_{\text{B}}\right)}&\leq\sigma_{\text{SBS}}^2\sqrt{\frac{z}{\tau f_s}\left(2P_{\text{PBS}}\text{Tr}\left(\hat{\boldsymbol{E}}\boldsymbol{\bar{\Phi}}[j-1]\right)/\sigma_{\text{SBS}}^2+N_{\text{B}}\right)}\\
		&+\frac{zP_{\text{PBS}}\text{Tr}\left(\hat{\boldsymbol{E}}\left(\boldsymbol{\bar{\Phi}}-\boldsymbol{\bar{\Phi}}[j-1]\right)\right)}{\tau f_s\sqrt{\frac{z}{\tau f_s}\left(2P_{\text{PBS}}\text{Tr}\left(\hat{\boldsymbol{E}}\boldsymbol{\bar{\Phi}}[j-1]\right)/\sigma_{\text{SBS}}^2+N_{\text{B}}\right)}},
	\end{aligned}
\end{equation}
where $j$ denotes the number of SCA iteration.

Dinkelbach's transform can also be used to transform the constraints in (\ref{eq42b}) as
\begin{equation}
	\label{eq56}
	\begin{aligned}
		&a_{1,m}\text{Tr}\left(\boldsymbol{\overline{G}}_{m,m}\boldsymbol{\bar{\Phi}}\right)\!-\!\overline{y}_{1,m}\left(\sum\limits_{i=M+1}^{M+K}\text{Tr}\left(\boldsymbol{\overline{G}}_{m,i}\boldsymbol{\bar{\Phi}}\right)\!+\!\sigma_{\text{MS}}^2\right)\\
		&+c_m\left(1-P_d\left(\boldsymbol{\bar{\Phi}}\right)\right)\text{Tr}\left(\boldsymbol{\overline{G}}_{m,m}\boldsymbol{\bar{\Phi}}\right)-\overline{y}_{2,m}\left(\sum\limits_{i=M+1}^{M+K}\text{Tr}\left(\boldsymbol{\overline{G}}_{m,i}\boldsymbol{\bar{\Phi}}\right)+d_m\right)\geq t,\forall m\in\mathcal{M},
	\end{aligned}
\end{equation}
where $\overline{y}_{1,m}$ and $\overline{y}_{2,m}$ are updated by
\begin{equation}
	\label{eq57}
	\begin{aligned}
		&\overline{y}_{1,m}[i+1]=\frac{a_{1,m}\text{Tr}\left(\boldsymbol{\overline{G}}_{m,m}\boldsymbol{\bar{\Phi}}[i]\right)}{\sum\limits_{i=M+1}^{M+K}\text{Tr}\left(\boldsymbol{\overline{G}}_{m,i}\boldsymbol{\bar{\Phi}}[i]\right)+\sigma_{\text{MS}}^2},\\
		&\overline{y}_{2,m}[i+1]=\frac{c_m\left(1-P_d\left(\boldsymbol{\bar{\Phi}}[i]\right)\right)\text{Tr}\left(\boldsymbol{\overline{G}}_{m,m}\boldsymbol{\bar{\Phi}}[i]\right)}{\sum\limits_{i=M+1}^{M+K}\text{Tr}\left(\boldsymbol{\overline{G}}_{m,i}\boldsymbol{\bar{\Phi}}[i]\right)+d_m[i]},
	\end{aligned}
\end{equation}
where $i$ denotes the number of Dinkelbach's transform iteration. Similar with Theorem 1, the problem after Dinkelbach's transform is equivalent to (P4) and is guaranteed to convergent.

The SCA method is then used to approximate the constraints of the communication rate of SUEs. The constraint of the $k$-th SUE is rewritten as
\begin{equation}
	\label{eq58}
	\begin{aligned}
		&b\log_2\left(\sum\limits_{i=1}^{K}\text{Tr}\left(\boldsymbol{E}_{k,i}\boldsymbol{\bar{\Phi}}^*\right)+\sigma_{\text{SU}}^2\right)-b\log_2\left(\sum\limits_{i=1,i\neq k}^{K}\text{Tr}\left(\boldsymbol{E}_{k,i}\boldsymbol{\bar{\Phi}}^*\right)+\sigma_{\text{SU}}^2\right)\\
		&+\widetilde{c}\left(\frac{1}{12}\exp\left\{-\frac{z}{2}\right\}+\frac{1}{4}\exp\left\{-\frac{2z}{3}\right\}\right)\log_2\left(\sum\limits_{i=1}^{K}\text{Tr}\left(\boldsymbol{E}_{k,i}\boldsymbol{\bar{\Phi}}^*\right)+\overline{d}_k\right)\\
		&-\widetilde{c}\left(\frac{1}{12}\exp\left\{-\frac{z}{2}\right\}+\frac{1}{4}\exp\left\{-\frac{2z}{3}\right\}\right)\log_2\left(\sum\limits_{i=1,i\neq k}^{K}\text{Tr}\left(\boldsymbol{E}_{k,i}\boldsymbol{\bar{\Phi}}^*\right)+\overline{d}_k\right)\geq r_k.
	\end{aligned}
\end{equation}
The non-convex terms can be approximated using the first order Taylor expansion, which are expressed as
\begin{equation}
	\label{eq59}
	\begin{aligned}
		\log_2\left(\sum\limits_{i=1,i\neq k}^{K}\text{Tr}\left(\boldsymbol{E}_{k,i}\boldsymbol{\bar{\Phi}}^*\right)+\sigma_{\text{SU}}^2\right)&\leq\log_2\left(\sum\limits_{i=1,i\neq k}^{K}\text{Tr}\left(\boldsymbol{E}_{k,i}\boldsymbol{\bar{\Phi}}^*[j-1]\right)+\sigma_{\text{SU}}^2\right)\\
		&+\frac{\sum\limits_{i=1,i\neq k}^{K}\text{Tr}\left(\boldsymbol{E}_{k,i}\left(\boldsymbol{\bar{\Phi}}^*-\boldsymbol{\bar{\Phi}}^*[j-1]\right)\right)}{\ln2\left(\sum\limits_{i=1,i\neq k}^{K}\text{Tr}\left(\boldsymbol{E}_{k,i}\boldsymbol{\bar{\Phi}}^*[j-1]\right)+\sigma_{\text{SU}}^2\right)}
	\end{aligned}
\end{equation}
and
\begin{equation}
	\label{eq60}
	\begin{aligned}
		\log_2\left(\sum\limits_{i=1,i\neq k}^{K}\text{Tr}\left(\boldsymbol{E}_{k,i}\boldsymbol{\bar{\Phi}}^*\right)+\overline{d}_k\right)&\leq\log_2\left(\sum\limits_{i=1,i\neq k}^{K}\text{Tr}\left(\boldsymbol{E}_{k,i}\boldsymbol{\bar{\Phi}}^*[j-1]\right)+\overline{d}_k\right)\\
		&+\frac{\left(P_{\text{PBS}}+1\right)\sum\limits_{i=1,i\neq k}^{K}\text{Tr}\left(\boldsymbol{E}_{k,i}\left(\boldsymbol{\bar{\Phi}}^*-\boldsymbol{\bar{\Phi}}^*[j-1]\right)\right)}{\ln2\left(\sum\limits_{i=1,i\neq k}^{K}\text{Tr}\left(\boldsymbol{E}_{k,i}\boldsymbol{\bar{\Phi}}^*[j-1]\right)+\overline{d}_k[j-1]\right)},
	\end{aligned}
\end{equation}
where $j$ denotes the number of SCA iteration.

The alternating optimization (AO) method can be used to solve the joint optimization problem on $z$ and $\boldsymbol{\bar{\Phi}}$. All constraints in (P4) except the rank one constraint are convex when we keep the auxiliary variable $z$ fixed. The SDR method can be applied to relax the rank one constraint. (P4) is reformulated as
\begin{subequations}
	\label{eq61}
	\begin{align}
		\text{(P5)}\mathop{\text{max}}\limits_{t,\boldsymbol{\bar{\Phi}}}&\ t \label{eq61a} \tag{61a} \\
		\text{s.t.}& \ \text{(\ref{eq54})-(\ref{eq56}), (\ref{eq58})-(\ref{eq60}), (\ref{eq42g}), (\ref{eq42h})}.
	\end{align}
\end{subequations}
The above problem is a standard SDP problem, which can be solve by CVX tools. The Gaussian randomization \cite{so2007approximating} can ensure a rank one solution. We also apply an exhaustive search method to obtain $z$ since $z$ is a one-dimensional variable. The scheme to solve the RIS beamforming problem is presented in \textbf{Algorithm 2}. The overall BCD-based algorithm for designing the joint beamforming is presented in \textbf{Algorithm 3}.
\begin{algorithm}[t]
	\caption{Dinkelbach’s transform and SCA methods for solving (P4).}\label{alg:alg2}
	\small
	\begin{algorithmic}
		\STATE 
		\STATE 1. $ \textbf{Inputs: }$$\tau$, $\boldsymbol{W}$, $N_\text{B}$, $N_\text{R}$, $N_\text{M}$, $M$, $K$, $L$, $\Pr\left\{\mathcal{H}_0\right\}$, $\Pr\left\{\mathcal{H}_1\right\}$, $T$, $r_k$, $P_{\text{PBS}}$, $\gamma_{l,0}$, $\lambda_m$, $\sigma_{\text{MS}}^2$, $\sigma_{\text{SU}}^2$, $\sigma_{\text{SBS}}^2$, $i_{\text{max}}$, $j_{\text{max}}$, $k_{\text{max}}$, $\varepsilon$, all channel matrices and vectors. 
		\STATE 2. $ \textbf{Outputs: }$Beamforming matrix $\boldsymbol{\Phi}$ for (P4).
		\STATE 3. $ \textbf{Initialization: }$Set $t=0$, solve (P5) and perform Gaussian randomization to obtain $\boldsymbol{\Phi}_{\text{DT}}[0]$, which represent the initial RIS beamforming matrix of Dinkelbach's transform, and calculate the sensing SINR as $O_{\text{DT}}[0]$. Set $O_{\text{DT}}[1]=+\inf$. Set $i,j,k=1$.
		\WHILE {$i\!\leq\! i_{\text{max}}$ and $\left\vert O_{\text{DT}}[i]-O_{\text{DT}}[i-1]\right\vert/\left\vert O_{\text{DT}}[i-1]\right\vert\!\geq\!\varepsilon$}
		\STATE 4.Calculate $\overline{y}_{1,m}[i]$ and $\overline{y}_{2,m}[i]$ according to (\ref{eq57}).
		\STATE 5.Set $\boldsymbol{\Phi}_{\text{AO}}[0]=\boldsymbol{\Phi}_{\text{DT}}[i-1]$, $O_{\text{AO}}[0]=O_{\text{DT}}[i-1]$, and $O_{\text{AO}}[1]=+\inf$. Calculate $z_{\text{AO}}[0]$ using $\boldsymbol{\Phi}_{\text{AO}}[0]$ and (\ref{eq52}).
		\WHILE {$j\!\leq\!j_{\text{max}}$ and $\left\vert O_{\text{AO}}[j]\!-\!O_{\text{AO}}[j\!-\!1]\right\vert\!/\!\left\vert O_{\text{AO}}[j\!-\!1]\right\vert\!\!\geq\!\varepsilon$}
		\STATE 6.Set $\boldsymbol{\Phi}_{\text{SCA}}[0]=\boldsymbol{\Phi}_{\text{AO}}[i-1]$, $O_{\text{SCA}}[0]=O_{\text{AO}}[i-1]$, and $O_{\text{SCA}}[1]=+\inf$.
		\WHILE{$\left\vert O_{\text{SCA}}[k]-O_{\text{SCA}}[k-1]\right\vert/\left\vert O_{\text{SCA}}[k-1]\right\vert\geq\varepsilon$ and $k\leq k_{\text{max}}$}
		\STATE 7.Solve (P5) and perform Gaussian randomization to obtain $\boldsymbol{\Phi}_{\text{SCA}}[k]$ and $O_{\text{SCA}}[k]$.
		\ENDWHILE
		\STATE 8.Set $\boldsymbol{\Phi}_{\text{AO}}[j]=\boldsymbol{\Phi}_{\text{SCA}}[k]$.
		\STATE 9. Use the exhaustive method to obtain $z[j]$. Calculate the sensing SINR as $O_{\text{AO}}[j]$.
		\ENDWHILE
		\STATE 10.Set $\boldsymbol{\Phi}_{\text{DT}}[i]=\boldsymbol{\Phi}_{\text{AO}}[j]$ and $O_{\text{DT}}[i]=O_{\text{AO}}[j]$.
		\ENDWHILE
		\STATE 11.Set $\boldsymbol{\Phi}=\boldsymbol{\Phi}_{\text{DT}}[i]$.
	\end{algorithmic}
	\label{alg2}
\end{algorithm}
\begin{algorithm}[t]
	\caption{BCD-based algorithm for solving (P1).}\label{alg:alg3}
	\small
	\begin{algorithmic}
		\STATE 
		\STATE 1. $ \textbf{Inputs:}$ $N_\text{B}$, $N_\text{R}$, $N_\text{M}$, $M$, $K$, $L$, $\Pr\left\{\mathcal{H}_0\right\}$, $\Pr\left\{\mathcal{H}_1\right\}$, $T$, $r_k$, $P_{\text{SBS}}$,  $P_{\text{PBS}}$, $\gamma_{l,0}$, $\lambda_m$, $\sigma_{\text{MS}}^2$, $\sigma_{\text{SU}}^2$, $\sigma_{\text{SBS}}^2$, $i_{\text{max}}$, $\varepsilon$, all channel matrices and vectors. 
		\STATE 2. $ \textbf{Outputs: }$Beamforming matrices $\boldsymbol{W}$ and $\boldsymbol{\Phi}$ for (P2) and (P4).
		\STATE 3. $ \textbf{Initialization: }$Perform the initialization of Algorithms 1 and 2 to obtain $\boldsymbol{W}[0]$ and $\boldsymbol{\Phi}[0]$. Randomize $\tau[0]$ and calculate the sensing SINR as $O[0]$. Set $O[1]=+\inf$ and $i=1$.
		\WHILE {$i\leq i_{\text{max}}$ and $\frac{\left\vert O[i]-O[i-1]\right\vert}{\left\vert O[i-1]\right\vert}\geq\varepsilon$}
		\STATE 4.Fix $\boldsymbol{W}[i-1]$ and $\boldsymbol{\Phi}[i-1]$, use the exhaustive search method in Section \uppercase\expandafter{\romannumeral3}.A to obtain $\tau[i]$.
		\STATE 5.Fix $\tau[i]$ and $\boldsymbol{\Phi}[i-1]$, perform Algorithm 1 to obtain $\boldsymbol{W}[i]$.
		\STATE 6.Fix $\tau[i]$ and $\boldsymbol{W}[i]$, perform Algorithm 2 to obtain $\boldsymbol{\Phi}[i]$.
		\ENDWHILE
		\STATE 7.Set $\tau=\tau[i]$, $\boldsymbol{W}=\boldsymbol{W}[i]$ and $\boldsymbol{\Phi}=\boldsymbol{\Phi}[i]$.
	\end{algorithmic}
	\label{alg3}
\end{algorithm}

\section{Complexity and Convergence Analysis}
We focus on the complexity of the BS beamforming and the RIS beamforming since they dominate the complexity.

\subsection{Complexity Analysis}
The SDR problems with $m$ constraints and the dimension of matrix in each constraint equals $n\times n$ can be solved with a worst-case complexity of $\mathcal{O}\left(\log\left(1/\eta\right)\max\left\{m,n\right\}^4n^{0.5}\right)$, where $\eta>0$ is the solution accuracy \cite{luo2010semidefinite}. Therefore, the complexity for solving the problem of BS beamforming is
\begin{equation}
	\label{eq62}
	\begin{aligned}
		\mathcal{O}_{\text{BS}}\!=\!\mathcal{O}\Big(\!\log\left(1/\eta\right)N_{\text{DT}}^{\text{BS}}N_{\text{SCA}}^{\text{BS}}\max\left\{N_{\text{B}},2(M\!+\!K)\!+\!L\!+\!1\right\}^4\left(2M+2K+L+1\right)^{0.5}\Big),
	\end{aligned}
\end{equation}
where $N_{\text{DT}}^{\text{BS}}$ and $N_{\text{SCA}}^{\text{BS}}$ represent the number of iterations of Dinkelbach's transform and the SCA. The complexity for solving the problem of RIS beamforming is
\begin{equation}
	\label{eq63}
	\begin{aligned}
		\mathcal{O}_{\text{RIS}}=\mathcal{O}\Big(\log\left(1/\eta\right)N_{\text{DT}}^{\text{RIS}}N_{\text{AO}}^{\text{RIS}}N_{\text{SCA}}^{\text{RIS}}\left(N_{\text{R}}+M+K+L+3\right)^4\left(N_{\text{R}}+1\right)^{0.5}\Big),
	\end{aligned}
\end{equation}
where $N_{\text{DT}}^{\text{RIS}}$, $N_{\text{AO}}^{\text{RIS}}$, and $N_{\text{SCA}}^{\text{RIS}}$ represent the number of iterations of the Dinkelbach's transform, the AO, and the SCA, respectively. The overall complexity of the BCD-based complexity is $\mathcal{O}_{\text{BCD}}=N_{\text{BCD}}\left(\mathcal{O}_{\text{BS}}+\mathcal{O}_{\text{RIS}}\right)$, where $N_{\text{BCD}}$ is the number of iteration of the BCD-based algorithm. We will evaluate the number of iteration of all iterative algorithms through simulations in the next section.

\subsection{Convergence Analysis}
The convergence of Dinkelbach's transform is proofed in Theorem 1. The SCA method surrogates a lower bound for the communication rate and iteratively approaches the real communication rate; it also converges. Moreover, the convergence performance of the AO algorithm and the BCD algorithm will be demonstrated by simulations in the next section.

\section{Simulation Results}
\begin{table}[!t]
	\centering
	\caption{Parameter values in our simulations.}
	\label{table:1}
	\begin{tabular}{|l| l| l|}
		\hline
		$\sigma^2$ & Noise variance & -60 dBm \\ 
		\hline
		$\Pr\left\{\mathcal{H}_0\right\}$ & Probability of $\mathcal{H}_0$ & 0.8 \\
		\hline
		$\Pr\left\{\mathcal{H}_1\right\}$ & Probability of $\mathcal{H}_1$ & 0.2 \\
		\hline
		$f_c$ & Carrier center frequency & 3 GHz \\
		\hline
		$f_s$ & Sampling frequency of the SBS & 6 MHz \\
		\hline
		$T$ & Downlink time duration & 10 ms\\
		\hline
		$\lambda_m$ & weighted factor of MS & 0.5 \\
		\hline
		$P_{\text{PBS}}$ & Transmit power of the PBS & 35 dBm \\
		\hline
		$r_k$ & Communication rate threshold of SUE & 2 \\
		\hline
		$\gamma_{l,0}$ & Interference threshold of the $l$-th PUE & -10 dBm \\
		\hline
		$P_{d,0}$ & Threshold of the detection probability & 0.9 \\
		\hline
		$P_{f,0}$ & Threshold of the false alarm probability & 0.1 \\
		\hline
		$\varepsilon$ & Error tolerance of iterative algorithms & 0.01 \\
		\hline
	\end{tabular}
\end{table}

We provide simulation results in this section to verify the effectiveness of the proposed algorithms. It is assumed that the SBS is located at (0m, 50m); the PBS is located at (0m, -50m); the cluster centre of the MSs is located at (70m, 30m); the cluster centre of the PUEs is located at (70m, -30m); the cluster centre of the SUEs is located at (25m, 50m). Since the secondary transmissions occur opportunistically, deploying the RIS near to the secondary network in advance is difficult. We will analyze the impact of the RIS location on the system performance. All channels associated with the RIS, i.e., $\boldsymbol{h}_{\text{P2R}}$, $\boldsymbol{H}_{\text{S2R}}$, $\boldsymbol{H}_{\text{R2MS},m}$, $\boldsymbol{h}_{\text{R2SU},k}$, and $\boldsymbol{h}_{\text{R2PU},l}$, are assumed to follow the Rician fading channel model and to be dominated by the line-of-sight (LoS) components. Their path loss factors are both equal to 2.2. All direct channels, i.e., $\boldsymbol{h}_{\text{P2S}}$, $\boldsymbol{h}_{\text{P2MS},m}$, $\boldsymbol{h}_{\text{P2SU},k}$, $\boldsymbol{H}_{\text{S2MS},m}$, $\boldsymbol{h}_{\text{S2SU},k}$, and $\boldsymbol{h}_{\text{S2PU},l}$, are assumed to follow Rayleigh fading channel model. Their path loss factors are both equal to 4. If not otherwise speciﬁed, the following simulations consider a RIS-enhanced cognitive ISAC system with two MSs, two SUEs, and two PUEs. The number of the SBS antennas, the MS antennas, and the RIS elements are equal to 16, 6, and 32, respectively. The transmit power of SBS is equal to 35 dBm. The parameter values of our simulations are listed in Table \uppercase\expandafter{\romannumeral1}.

\begin{figure}[!t] %hb代表放在文章底部，%ht为放在文章顶部
	\centering  %图片全局居中
	\subfigure[The sensing SINR versus the transmit power of the SBS.]{
		\includegraphics[width=0.45\linewidth]{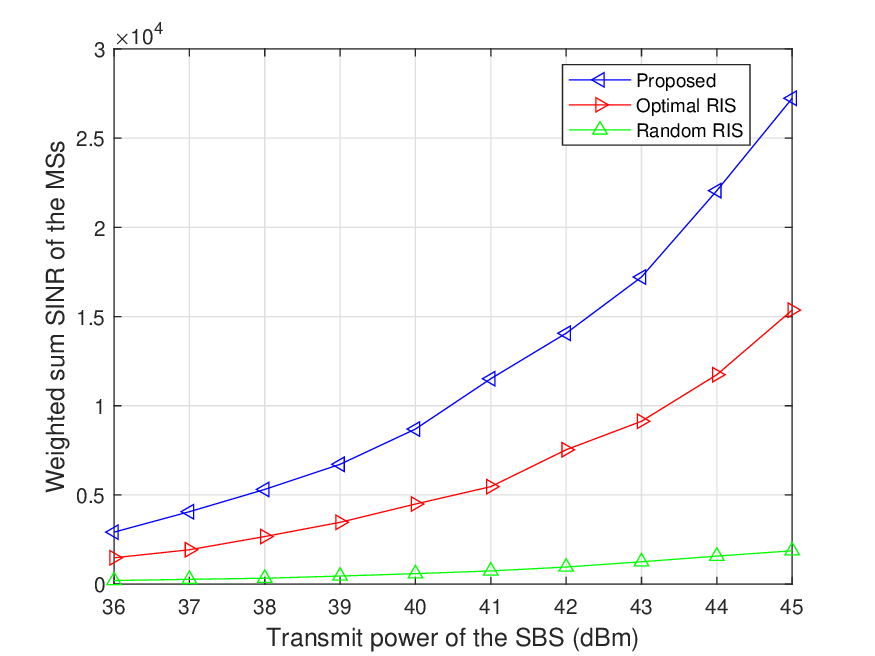}}
	\subfigure[The weighted PEB versus the transmit power of the SBS.]{
		\includegraphics[width=0.45\linewidth]{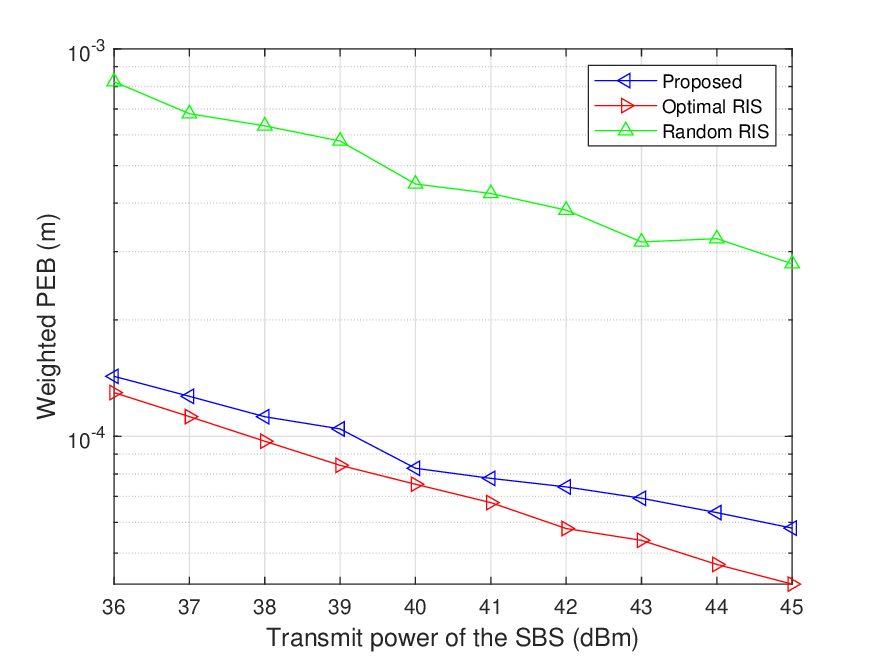}}
	\caption{The Sensing SINR and PEB versus the transmit power.}
	\label{fig2}
\end{figure}

\begin{figure}[!t] %hb代表放在文章底部，%ht为放在文章顶部
	\centering  %图片全局居中
	\subfigure[The sensing SINR versus the number of RIS elements.]{
		\includegraphics[width=0.45\linewidth]{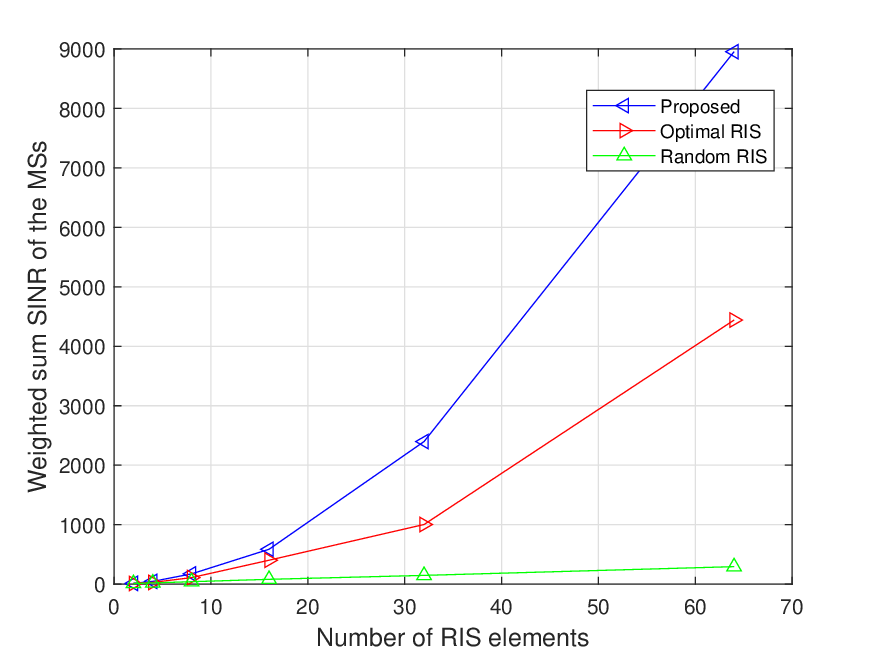}}
	\subfigure[The weighted PEB versus the number of RIS elements.]{
		\includegraphics[width=0.45\linewidth]{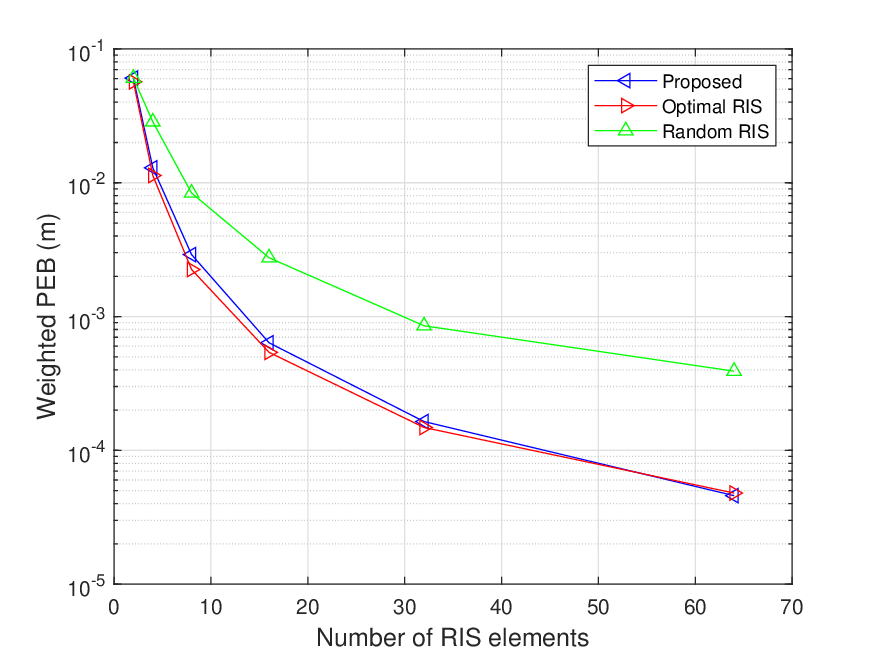}}
	\caption{The Sensing SINR and PEB versus the number of RIS elements.}
	\label{fig3}
\end{figure}

\begin{figure}[!t]
	\centering
	\subfigure[The sensing SINR versus the number of MS antennas.]{
		\includegraphics[width=0.45\linewidth]{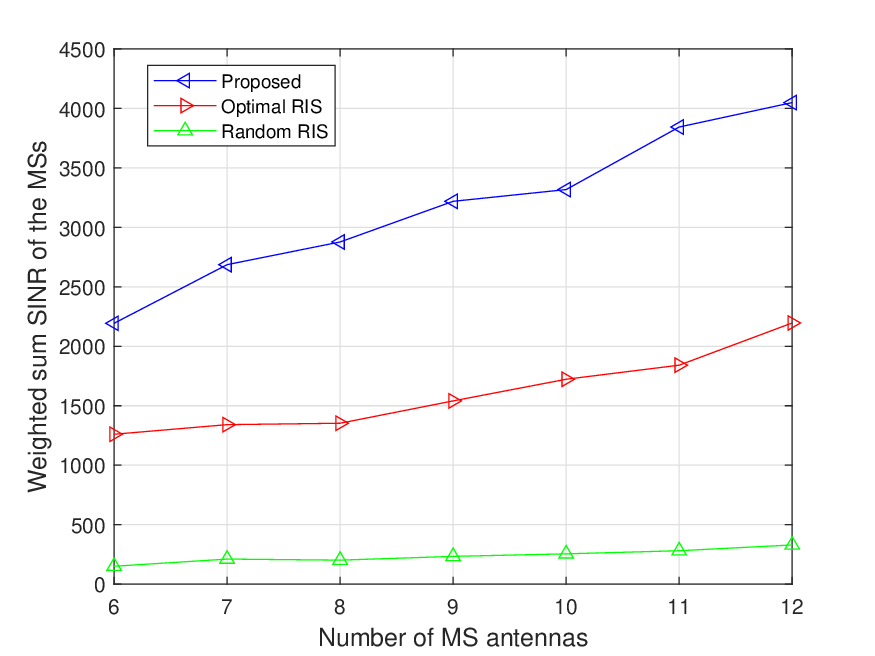}}
	\subfigure[The weighted PEB versus the number of MS antennas.]{
		\includegraphics[width=0.45\linewidth]{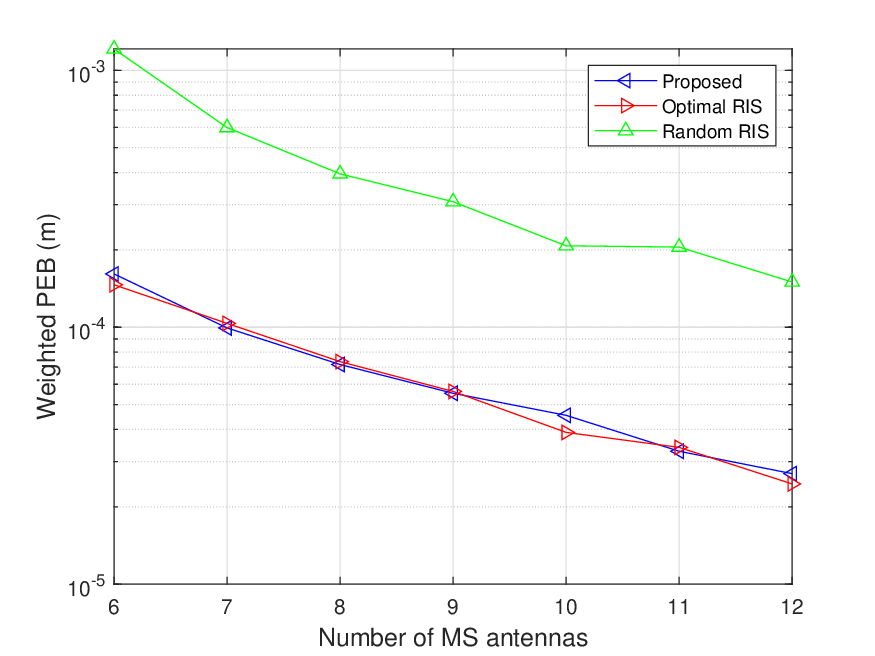}}
	\caption{The Sensing SINR and PEB versus the number of MS antennas.}
	\label{fig4}
\end{figure}

\begin{figure}[!t]
	\centering
	\includegraphics[width=0.8\textwidth]{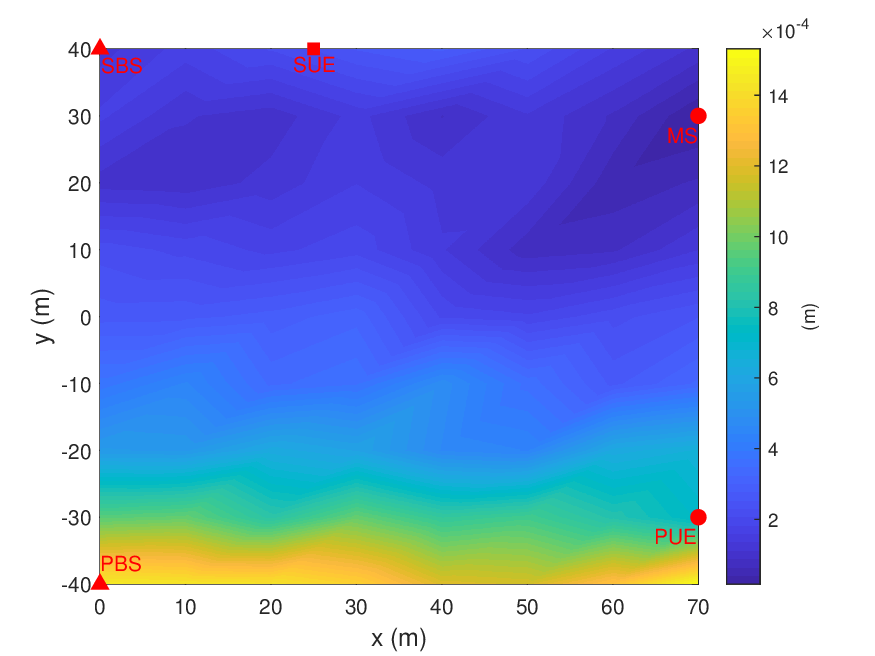}
	\caption{The weighted PEB versus the location of RIS.}
	\label{fig5}
\end{figure}

\begin{figure}[!t]
	\centering
	\includegraphics[width=0.8\textwidth]{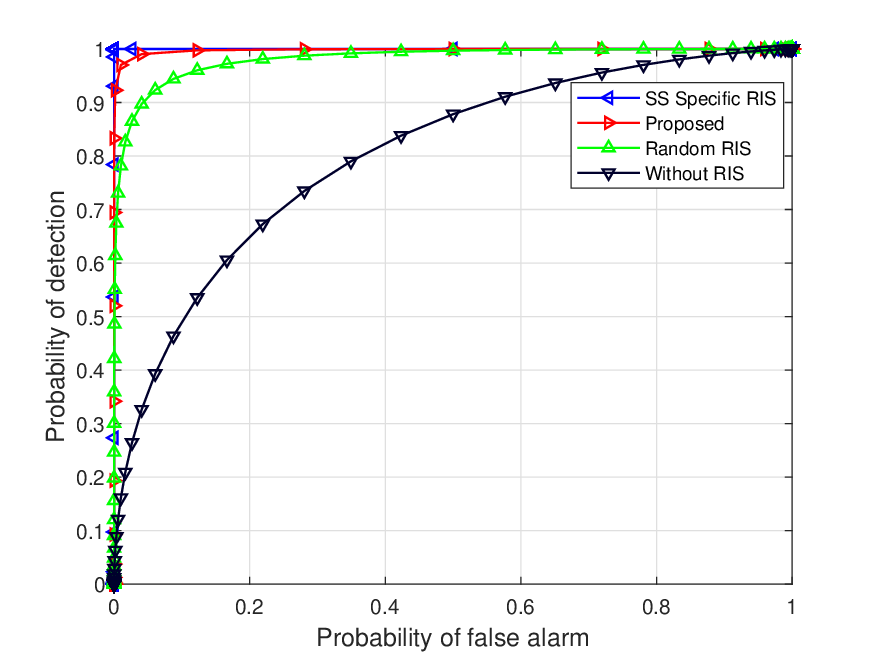}
	\caption{Complementary ROC curve for the considered RIS-enhanced cognitive ISAC system.}
	\label{fig6}
\end{figure}

\begin{figure}[!t]
	\centering
	\includegraphics[width=0.8\textwidth]{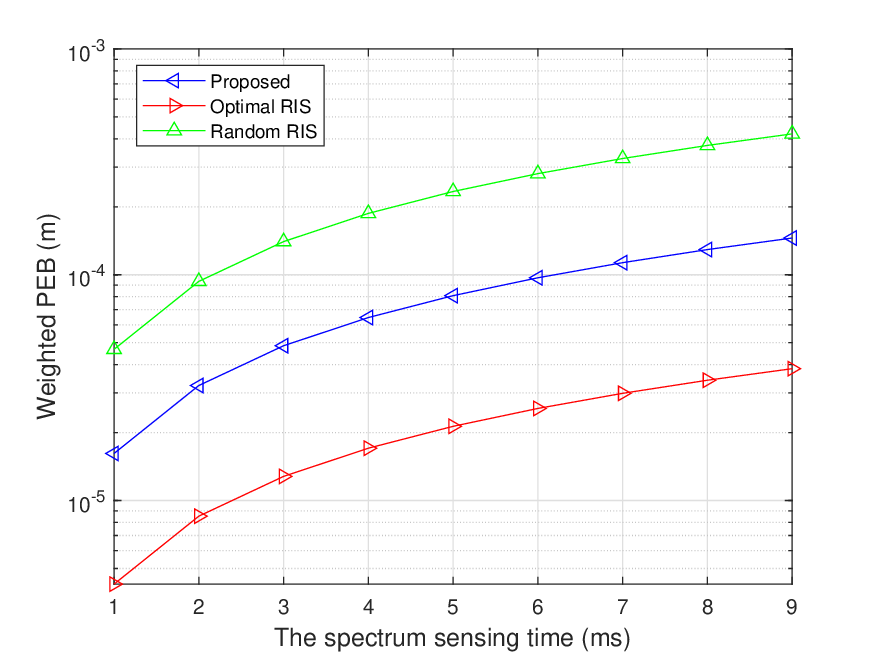}
	\caption{The weighted PEB versus the spectrum sensing time.}
	\label{fig7}
\end{figure}

\begin{figure}[!t]
	\centering
	\includegraphics[width=0.8\textwidth]{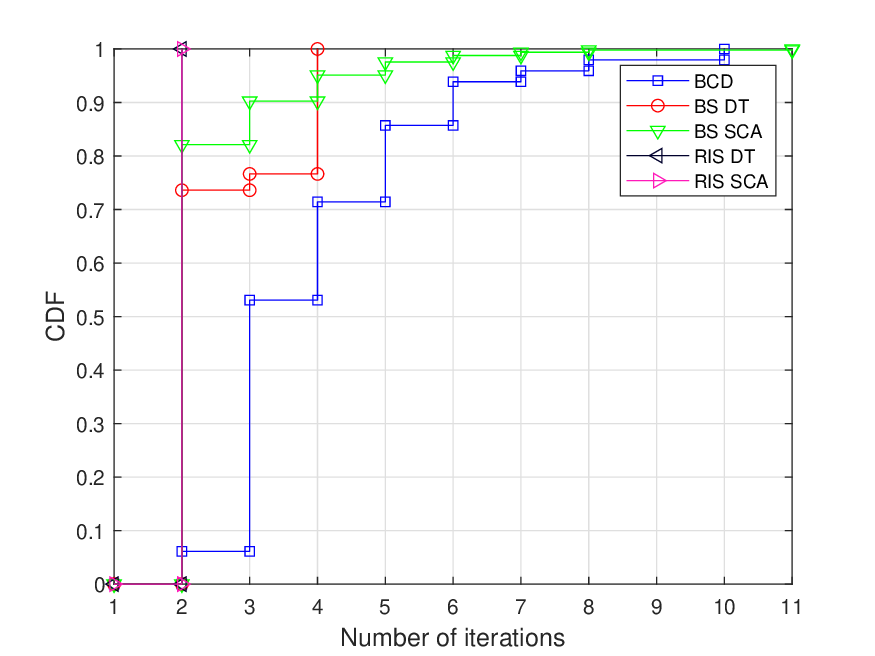}
	\caption{The cumulative distribution function (CDF) of the number of iterations.}
	\label{fig8}
\end{figure}

From Fig. \ref{fig2} to Fig. \ref{fig4}, ``Optimal RIS'' refers to using the proposed Algorithm 1 to obtain the BS beamforming matrix and designing the RIS beamforming matrix with a heuristic algorithm to minimize the weighted PEB directly; ``Random RIS'' refers to using the proposed Algorithm 1 to obtain the BS beamforming matrix and randomly obtaining the RIS beamforming matrix. Figure \ref{fig2} (a) shows that as the transmit power of the SBS increases, the weighted sum SINRs of MSs obtained by all schemes increase. The SINR obtained by the proposed scheme is larger than those of the optimal RIS and random RIS since the optimal RIS is not specifically used to maximize the weighted sum SINR. Moreover, in Fig. \ref{fig2} (b), it is observed that the weighted PEBs obtained by all schemes decrease as the transmit power of the SBS increases. Optimizing for the weighted sum SINR leads to the weighted PEB consistently decreasing with the increase of the weighted sum SINR. The gap between the proposed scheme and the optimal RIS is small and stable. This validates the effectiveness of the proposed scheme in improving localization accuracy while satisfying the SS and communication requirements. 

Figure \ref{fig3} (a) shows that as the number of RIS elements increases, the weighted sum SINRs of MSs obtained by all schemes increase. This is because the power gain provided by RIS becomes large when the number of RIS elements increases. The SINR obtained by the proposed scheme is larger than the optimal RIS and random RIS. In Fig. \ref{fig3} (b), it is shown that the weighted PEBs obtained by all schemes decrease as the number of RIS elements increases. The gap between the proposed scheme and the optimal RIS is also small and stable. 

The performance of the proposed scheme is further validated by Fig. \ref{fig4} (a) and (b), where the proposed scheme can also obtain a good weighted PEB compared to the optimal RIS when we increase the number of MS antennas.

We further evaluate the impact of the RIS location on the weighted PEB by Fig. \ref{fig5}. It is shown that the proposed scheme can obtain a lower weighted PEB when deploying the RIS near the SBS or the MS, whereas the proposed scheme yields a higher weighted PEB when the RIS is deployed far from both the SBS and the MS. The above phenomena can be attributed to the channel gains of the RIS cascaded channels, which are inversely proportional to the product of the distances in the cascaded channels. Therefore, we should carefully deploy RISs in CR networks, as deploying a RIS at an arbitrary location does not necessarily guarantee performance gain.

The performance of the RIS-enhanced SS is validated through Fig. \ref{fig6}, where ``SS Specific RIS'' refers to the RIS is only used to enhance the SS without considering the communication and the localization performance. Figure \ref{fig6} demonstrates that, for a fixed probability of false alarm, the proposed scheme achieves a higher detection probability than the random RIS scheme and the without RIS scheme and slightly lower than that of the SS specific RIS. This indicates that although the proposed scheme is not specifically designed to enhance the SS, it can still achieve satisfactory SS performance. 

Figure \ref{fig7} shows that as the SS time increases, the weighted PEBs of all schemes gradually increase. This is because the time used for transmitting the ISAC signals decreases as the SS time increases, and the weighted PEB only depends on the time of transmitting the ISAC signals and the SNR of the received signals at the MS.

Figure \ref{fig8} shows the convergence performance of the proposed algorithms. All iterative algorithms converge within 11 iterations. It is noted that the initial values of the SCA methods can influence their convergence performance. We can obtain the initial values of the SCA methods by directly maximizing the communication rate without considering the PEB performance of the MS.

\section{Conclusion}
We have investigated the intersections between the emerging technologies of 6G, i.e., the RIS and the ISAC, and the CR to further enhance the system performance of CR and ISAC. An optimization problem is formulated to maximize the SINRs of the MSs by designing the SS time, the BS beamforming, and the RIS beamforming. The formulated problem is also guaranteed to satisfy the SS and the communication requirements of the considered RIS-enhanced cognitive ISAC system. It can be solved by the proposed BCD algorithm, which iteratively solves the sub-problems of designing the SS time, the BS beamforming, and the RIS beamforming, where the Dinkelbach's transform and the SCA are used to transform the non-convex functions into convex ones. Simulation results demonstrate that the gap between the PEB obtained by the proposed scheme and that of the optimal RIS scheme is small, and the proposed scheme has a good convergence performance. Hence, the proposed scheme is efficient in improving the accuracy of localizing MSs and the accuracy of the REM. Extending studies will likely be required to investigate the RIS deployment since the secondary transmissions occur opportunistically, and the performance of the RIS-enhanced cognitive ISAC systems highly depends on the RIS location.

\section*{Appendix A\\Proof of Lemma 1}
The noise-less and interference-less part of the received signals at the $m$-th MS can be expressed as
\begin{equation}
	\label{a1}
	\begin{aligned}
	\overline{\boldsymbol{y}}_{\text{MS},m}&=\left(\boldsymbol{H}_{\text{R2MS},m}^T\boldsymbol{\Phi}\boldsymbol{H}_{\text{S2R}}^T+\boldsymbol{H}_{\text{S2MS},m}^T\right)\boldsymbol{w}_m\\
	&=\left(h_{\text{D}}\boldsymbol{b}\left(\theta_{\text{D}}\right)\boldsymbol{a}^T\!\left(\theta_{\text{D}}^{\prime}\right)\!+\!h_{\text{R}}\boldsymbol{b}\left(\theta_{\text{R}}\right)\boldsymbol{c}^T\!\left(\theta_{\text{R}}\right)\boldsymbol{\Phi}\boldsymbol{c}\left(\theta_{\text{R}}^{\prime}\right)\boldsymbol{a}^T\!\left(\theta_{\text{R}}^{\prime}\right)\right)\boldsymbol{w}_m.
	\end{aligned}
\end{equation}
Then, we can directly obtain each element of $\frac{\partial\left(\overline{\boldsymbol{y}}\right)_n}{\partial\boldsymbol{\xi}_{\text{c}}}$ as shown in (\ref{eq12})-(\ref{eq17}), where the partial terms can be calculated as 
\begin{equation}
	\label{a2}
	\begin{aligned}
	\frac{\partial\boldsymbol{b}(\theta_{\text{D}})}{\partial\theta_{\text{D}}}=&\left[1,e^{\left(-j\pi\sin\theta_{\text{D}}\right)},\cdots,e^{\left(-j\pi(N_{\text{M}}-1)\sin\theta_{\text{D}}\right)}\right]^T\\
	&\odot\left[0,-j\pi\cos\theta_{\text{D}},\cdots,-j\pi(N_{\text{M}}\!-\!1)\cos\theta_{\text{D}}\right]^T,
	\end{aligned}
\end{equation}
and
\begin{equation}
	\label{a3}
	\begin{aligned}
		&\frac{\partial\boldsymbol{c}(\theta_{\text{R}})\boldsymbol{b}^T(\theta_{\text{R}})}{\partial\theta_{\text{R}}}=\frac{\partial\boldsymbol{c}(\theta_{\text{R}})}{\partial\theta_{\text{R}}}\boldsymbol{b}^T(\theta_{\text{R}})+\boldsymbol{c}(\theta_{\text{R}})\frac{\boldsymbol{b}^T(\theta_{\text{R}})}{\partial\theta_{\text{R}}}\\
		&=\left[1,e^{\left(-j\pi\sin\theta_{\text{R}}\right)},\cdots,e^{\left(-j\pi(N_{\text{R}}-1)\sin\theta_{\text{R}}\right)}\right]^T\odot\left[0,-j\pi\cos\theta_{\text{R}},\cdots,-j\pi(N_{\text{R}}-1)\cos\theta_{\text{R}}\right]^T\boldsymbol{b}^T(\theta_{\text{R}})\\
		&+\boldsymbol{c}(\theta_{\text{R}})\left[1,e^{\left(-j\pi\sin\theta_{\text{R}}\right)},\cdots,e^{\left(-j\pi(N_{\text{M}}-1)\sin\theta_{\text{R}}\right)}\right]odot\left[0,-j\pi\cos\theta_{\text{R}},\cdots,-j\pi(N_{\text{M}}-1)\cos\theta_{\text{R}}\right].
	\end{aligned}
\end{equation}
Moreover, the Jacobian matrix $\boldsymbol{\Gamma}$ can be calculated as
\begin{equation}
	\label{a4}
	\boldsymbol{\Gamma}=\begin{bmatrix}
		\begin{matrix}
			\partial\theta_{\text{D}}/\partial x,& \partial\theta_{\text{D}}/\partial y,\\
			\partial\theta_{\text{R}}/\partial x,& \partial\theta_{\text{R}}/\partial y, 
		\end{matrix}&\boldsymbol{0}_{2\times4},\\
		\boldsymbol{0}_{2\times4},&\boldsymbol{\text{I}}_{4\times4}
	\end{bmatrix}.
\end{equation}
We can obtain the partial terms in (\ref{a4}) by $\theta_{\text{D}}=\arctan\left(\vert y-y_{\text{SBS}}\vert/\vert x-x_{\text{SBS}}\vert\right)$ and $\theta_{\text{R}}=\arctan\left(\vert y-y_{\text{RIS}}\vert/\vert x-x_{\text{RIS}}\vert\right)$. This completes the proof of Lemma 1.

\bibliographystyle{IEEEtran}
\bibliography{CR-based_ISAC_singleColumn}

\end{document}